\documentclass[11pt,a4paper]{article}
\usepackage{subfigure}
\usepackage{jheppub}

\newcommand{\be}{\begin{eqnarray}}
\newcommand{\ee}{\end{eqnarray}}

\newcommand{\bea}{\begin{eqnarray}}
\newcommand{\eea}{\end{eqnarray}}  
\newcommand{\nn}{\nonumber}
\newcommand{\Tr}{\textrm{Tr}}

\newcommand{\NN}{\mathcal{N}}
\newcommand{\la}{\lambda}
 \newcommand{\sfrac}[2]{\mbox{$\frac{#1}{#2}$}}
\newcommand{\ms}{\!-\!}
\newcommand{\ps}{\!+\!}
\newcommand{\tl}{\tilde\lambda}
\newcommand{\pint}{\int\!\!\!\!\!\!-}
\newcommand{\pintba}{\int^a_b\!\!\!\!\!\!\!\!\!-}
\newcommand{\Li}{\mbox{Li}}
\newcommand{\kc}{\kappa_{crit}}
\newcommand{\CC}{{\mathcal C}}

\def\eps{\epsilon}

\title{Phases of planar $5$-dimensional supersymmetric Chern-Simons theory}
\author[a]{Joseph A. Minahan,}
\author[a]{Anton Nedelin}
\affiliation[a]{Department of Physics and Astronomy,
     Uppsala university,\\
     Box 516,
     SE-75120 Uppsala,
     Sweden}
\emailAdd{joseph.minahan@physics.uu.se}
\emailAdd{anton.nedelin @physics.uu.se}
\abstract{In this paper we investigate the large-$N$ behavior of 5-dimensional $\NN=1$
 super Yang-Mills  with   a level $k$ Chern-Simons term and an adjoint hypermultiplet.  As in three-dimensional Chern-Simons theories, one must  choose an integration contour to completely define the theory.  Using localization, we reduce the path integral to a matrix model with a cubic action and compute its free energy in various scenarios.  In the limit of infinite Yang-Mills coupling and for particular choices of the contours,  we find that the free-energy scales as $N^{5/2}$    for  $U(N)$ gauge groups with large values of the Chern-Simons 't\,Hooft coupling, $\tl\equiv N/k$.   If we also set the hypermultiplet mass to zero, then this limit is a superconformal fixed point and the $N^{5/2}$ behavior parallels other  fixed points which have known supergravity duals.  We also demonstrate that $SU(N)$ gauge groups cannot have this  $N^{5/2}$ scaling for their free-energy.   At finite Yang-Mills coupling we establish the existence of a third order phase transition where the theory crosses over from the Yang-
Mills phase to the Chern-Simons phase.  The phase transition exists for any value of $\tl$, although the details differ between small and large values of $\tl$.  
 For pure Chern-Simons theories we  present evidence for a chain of phase transitions as $\tl$ is increased.  
 
 We also find the 
 expectation values for supersymmetric circular Wilson loops in these various scenarios and show that  the Chern-Simons term leads to different physical properties for fundamental and anti-fundamental  Wilson loops.  Different choices of the integration contours also lead to different properties for the loops.}
 
\arxivnumber{1408.2767}

\begin{document}

\maketitle

\section{Introduction and main results}

There  has been much 
interest in 5-dimensional supersymmetric gauge theories, in part because of their  relation to $6D$ $(2,0)$ superconformal field theories \cite{Witten:1995ex,Lambert:2010iw,Douglas:2010iu}.   Using  localization it is possible to compute the free-energies of $\NN=1$ and $\NN=2$ super Yang-Mills (SYM) on $S^5$ \cite{Kallen:2012cs,Hosomichi:2012ek,Kallen:2012va,Kim:2012av}. 
In particular, at the $\NN=2$ point in \cite{Kim:2012av} and more generally in \cite{Kallen:2012zn,Minahan:2013jwa} it was shown that the free-energy of $\NN=1$ SYM with an adjoint hypermultiplet behaves 
as $N^3$ in the planar limit at strong coupling, with the coefficient dependent on the hypermultiplet mass. The $N^3$ behavior is consistent with supergravity considerations, where one can show that the free-energy of  the 6D theory compactified on $S^5\times S^1$ also scales as $N^3$ \cite{Klebanov:1996un,Henningson:1998gx}.  One also finds $N^3$ behavior on squashed spheres, where the only difference  with the sphere  is an overall volume factor in the free-energy \cite{Qiu:2013pta}.

Other 5-dimensional  theories of interest  are superconformal fixed points
\cite{Seiberg:1996bd,Intriligator:1997pq}, which are  the infinite coupling limits of certain SYM theories.
The conformal fixed points  can be divided into three main classes. The first has super Yang-Mills 
with  exceptional gauge groups. We won't speak further about these here.

The second class is super Yang-Mills
 with a $USp(N)$ gauge group. These theories are interesting because they have  known holographic $AdS_6$ duals
\cite{Ferrara:1998gv,Brandhuber:1999np}. Recently, using a brane network construction this class of $USp(N)$ theories
were generalized to quiver theories and their $AdS_6$ duals \cite{Bergman:2012kr}. 
The gauge theories were studied using  localization  in \cite{Jafferis:2012iv} and
\cite{Assel:2012nf}. Here it was observed  that the free-energies  behave as $N^{5/2}$ and agree with the corresponding supergravity computation on the $AdS_6$ duals.

The third type of 5-dimensional superconformal theory, and the principle focus of this paper, is  a strong coupling limit of  $U(N)$ or $SU(N)$ SYM  with a Chern-Simons (CS) term in the action.   $U(N)$ and $SU(N)$ are the only groups that allow for a nontrivial CS action.  Consideration of a CS term is not just an idle exercise, as it can be generated by integrating out massive hypermultiplets in complex representations \cite{Seiberg:1996bd,Intriligator:1997pq,Minahan:2013jwa}.  The CS level $k$ is quantized, but we are interested in the case of large $k$, where one can define an 't\,Hooft parameter $\tilde\lambda={N}/{k}$, with $\tilde\lambda$ fixed in the large-$N$ limit, and essentially continuous.  One of the interesting issues we  observe here is that the large-$N$ behavior is significantly different between the $U(N)$ and $SU(N)$ theory because of the cubic nature of the action. In particular, with a proper choice of contour the  $U(N)$ free-energy  exhibits $N^{5/2}$
behavior at large $\tilde\lambda$,  analogous to the $USp(N)$ Yang-Mills result in \cite{Jafferis:2012iv}.  Such behavior does not seem possible for the
$SU(N)$  free-energy.   The $N^{5/2}$ dependence suggests the possible existence of an $AdS_6$ supergravity dual, although we presently do not know of one.  In fact,  there are other reasons to believe that  a supergravity dual might not exist, as we will explain later in the paper.

Another interesting issue is the interplay between  SYM and CS behavior.
In the large $N$ limit we should expect a sharp crossover between an SYM phase and a CS phase.  To investigate this crossover we consider having  finite 't\,Hooft parameters    $\tilde\lambda$ and $\lambda\equiv g_{YM}^2N/r$,  where $r$ is the $S^5$ radius, which leads to an action with   cubic and  quadratic pieces.  Here we will find a phase transition as the ratio 
\be\label{kappadef}
\kappa\equiv 8\pi^2 \frac{\tilde\lambda}{\lambda}
\ee
 reaches a critical value.   The critical values are different for $U(N)$ and $SU(N)$ and they are also different for small or large values of the 't\,Hooft parameters.  But in all cases the phase transition is third order.  

In order to investigate  the 5-dimensional SYM-CS theory on $S^5$ we will use the  localization results  in \cite{Kallen:2012cs,Kallen:2012va}.
Localization reduces the path integral  to a matrix integral, vastly simplifying the
computations. 
We then proceed to solve the  matrix model in the large-$N$ limit, both analytically and numerically. As we will see, at weak coupling our matrix model will 
simplify  to a cubic matrix model with a logarithmic potential between the eigenvalues, which was well studied  in the context
of $2D$ quantum gravity \cite{DiFrancesco:1993nw}.  In this case the model can be solved by saddle point \cite{Brezin:1977sv}, leading to a generic solution with a continuous distribution of   eigenvalues lying on two cuts.  Different solutions to the matrix model correspond to different choices of an integration contour \cite{Witten:2010cx,Felder:2004uy}, whose choice is necessary to completely define the theory.

At infinite $\lambda$, we have a ``pure CS" model, where the action is entirely cubic.   If we also have $\tilde\lambda\ll1$, which we call the pure CS model at weak coupling, then the model has a $Z_3$ symmetry in the complex plane and we can look for solutions that are $Z_3$-symmetric.  Indeed such a solution exists, which we refer to as the $Z_3$ solution, and is a  limit where the end of one of the two cuts of the general solution meets the side of the other cut.
There are also three distinct single-cut solutions, which break the $Z_3$ symmetry but are transformed into each other under the   $Z_3$ group \cite{Marino:2012zq}.     One of the solutions is real, in that the cut is invariant under complex conjugation, while the other two solutions are complex conjugates of each other.  The $Z_3$-symmetric solution is valid for both $U(N)$ and $SU(N)$ gauge groups, but the single-cut solutions only apply to  $U(N)$  and not $SU(N)$ since  the eigenvalues do not preserve the traceless condition for the scalar fields.

The free-energy is computable for the $Z_3$-symmetric and single-cut solutions, where in both situations
it scales as $N^2$.  The  free-energy is the same for the single-cut solutions because of the $Z_3$ symmetry,  but as we will show, it is higher  than the free-energy for the $Z_3$-symmetric solution.
 
 As we increase $\tilde\lambda$, the $Z_3$ symmetry is explicitly broken by the determinant factors in the matrix model.  Starting with  the complex single-cut solutions, half the eigenvalues migrate exponentially close to the positive real axis and extend out to  order $\tilde\lambda^{1/2}$, while the other half move toward the positive (negative) imaginary axis  and  extend out to the same order.   Moreover, the eigenvalues on the real axis are part of a single cut, but numerical and analytic evidence shows that those on the imaginary axis split into order $\tilde\lambda^{1/2}$ separate cuts, indicating the crossing of phase transitions as the coupling is increased.    The existence of the phase transitions likely complicates the search for supergravity duals, since they appear at large $\tl$ where a dual would also be found.  The real single-cut solution behaves significantly differently from the single-cut complex solutions.  Here the eigenvalues remain on a single cut of finite extent as $\tilde\lambda$ 
approaches infinity.
 
 If we  instead start with the $Z_3$ solution\footnote{We will continue to refer to this solution as the $Z_3$ solution, even though the $Z_3$ symmetry is explicitly broken away from weak coupling.}, then as we increase $\tl$  the number of real eigenvalues increases  from a third to a half the overall number and their profile closely approaches the profile  of the  complex solutions along the real line.    However, the complex eigenvalues should appear in conjugate pairs and distribute themselves equally toward the positive and negative imaginary axes.  We expect their profiles to look like the imaginary part of the combined complex solutions,  breaking into multi-cuts on both sides of the real line, but with half the density.
 
  The dominant contribution to the real part of the free energy comes from the eigenvalues on or near the real line, where one finds the approximate result
\be\label{strongFE}
\mathrm{Re}\left(
F_{\mathrm{strong}}\right)\approx -\frac{9\pi}{20}N^2 \tilde\lambda^{1/2}\,,
\ee
at the superconformal point with zero hypermultiplet mass.
Given the $N$ dependence of $\tilde\lambda$,  this gives the aforementioned $N^{5/2}$ behavior. The eigenvalues along the imaginary axis contribute  the same order to the imaginary part of the free-energy.  However,  the imaginary part cancels out for the $Z_3$ solution and (\ref{strongFE}) is the complete free-energy.   The real single-cut solution also has a real free-energy.  However, because the eigenvalues  have  finite extent as $\tl\to\infty$, the free-energy only scales as $N^2$ in this limit.

One can also compute the expectation value of a  Wilson loop around a great circle of 
the $S^5$ using localization \cite{Pestun:2007rz,Minahan:2013jwa}.  However, because 
the CS term breaks charge conjugation invariance, the Wilson loop in a fundamental 
representation differs from the Wilson loop in the anti-fundamental representation.  
At weak coupling, the difference is just  a sign for the log of the Wilson loop. 
However, at strong coupling the difference is more pronounced.   For the fundamental
representation we find that $\log(\langle W\rangle)\sim \sqrt{\tl}$, while it is relatively 
suppressed for the anti-fundamental representation.

Going back to the case with finite $\lambda$, we will argue that the single-cut solution below the phase
transition can continuously connect to a double-cut solution above the transition which has lower free-energy 
than the complex single-cut solutions.  The eigenvalue distribution of this double-cut solution is symmetric 
about the real axis and hence the free-energy is real.  As we approach the pure CS case at strong coupling, 
the free-energy  is the same as in (\ref{strongFE}).

 This paper is organized as follows: In section 2 we  briefly review  the matrix model  obtained by localization of the SYM-CS theory on $S^5$. In section 3 we  solve the pure CS matrix model 
 at weak coupling where it reduces to a purely cubic matrix model with a logarithmic interaction potential.
 In section 4 we  solve the strong coupling limit of the pure CS model using particular approximations 
 that we check with numerical solutions. In section 5 we calculate the Wilson loop expectation values for the different pure CS solutions.  In section 6 we generalize our results to  quiver theories.  In section 7 we consider the case of finite $\lambda$ and study the phase transitions at both weak and strong coupling.  In section 8 we offer some concluding remarks.  Various technical discussions are contained in the appendices.
 
 \section{ Matrix model for $\NN=1$ $5D$ Yang-Mills with Chern-Simons and matter}
 \label{S-general-matrix}

  In order to  study the properties of $5D$ CS theory with matter we will use results of supersymmetric localization  
  \cite{Kallen:2012cs,Kallen:2012va}. Localization reduces the $S^5$ partition function of $5D$ SYM with a CS term and a matter multiplet in the $\rm{R}$ representation  to the matrix integral
      \be
Z&=&\int\limits_{\rm Cartan} [d\phi]~e^{-  \frac{8\pi^3 r}{g_{YM}^2}  \text{Tr}(\phi^2)-\frac{\pi k}{3}\text{Tr}
(\phi^3)}  Z_{\rm 1-loop}^{\rm vect} (\phi)    Z_{\rm 1-loop}^{\rm hyper} (\phi) + \mathcal{O} (e^{-\frac{16 \pi^3 r}
{g_{YM}^2}})~,\label{vh1loop-intro}
\ee
 where the one-loop contributions are given by
\bea
 Z_{\rm 1-loop}^{\rm vect} (\phi) =\prod\limits_\beta\prod\limits_{t \neq 0}\left( t - \langle \beta,i\phi\rangle  
 \right)^{(1+\frac{3}{2}t+\frac{1}{2}t^2)}~,\label{vect1-loop-beg}
\eea
 and
\begin{equation}\label{main-form-det}
Z_{\rm 1-loop}^{\rm hyper} (\phi) = \prod\limits_\mu \prod\limits_{t}\left( t - \langle i\phi , 
\mu\rangle -i m+\frac{3}{2} \right)^{-(1+\frac{3}{2}t+\frac{1}{2}t^2)}~.
\end{equation}
 Here $\beta$ are the roots,  $\mu$ are  the weights in  $\rm{R}$, $r$  is the radius of $S^5$,  and $m=-i M r$ with
 $M$ being the mass of the hypermultiplet.
This matrix model was studied in detail in \cite{Kallen:2012zn,Minahan:2013jwa,Kim:2012qf} for  the planar limit
 of $SU(N)$ SYM theory, usually ignoring the CS term, and in \cite{Jafferis:2012iv,Assel:2012nf} for $5D$ 
 superconformal theories.  The most interesting behavior occurs when we have a single hypermultiplet in the adjoint representation.

In the large  $N$ limit the  matrix integral in (\ref{vh1loop-intro}) is dominated by the saddle point.  
The matrix integral (and thus the corresponding saddle point equations)
 takes the same form for either  $U(N)$ or $SU(N)$ gauge groups, but for $SU(N)$ the sum of the eigenvalues $\phi_i$ of the matrix $\phi$ is constrained to be zero.  For SYM with no CS term the solution automatically satisfies the constraint because of a $Z_2$ symmetry, hence there is little distinction between the two groups.  However,  a CS term breaks the $Z_2$ symmetry  and the constraint has to be enforced  using a Lagrange multiplier.
 
 If we consider the hypermultiplet  to be in the adjoint representation, then in the large $N$ limit for $U(N)$ or $SU(N)$
 the partition function (\ref{vh1loop-intro}) is  dominated by the saddle point satisfying the equations 
\begin{eqnarray}
\nonumber
\frac{\pi N}{\tilde\lambda}\left(\phi^2_i+2\kappa\phi_i-\mu\right)&=& \pi \sum\limits_{j\neq i}
\Bigg[\left(2- (\phi_i-\phi_j)^2\right)\coth(\pi(\phi_i-\phi_j))
\\
&&\qquad\qquad+\frac12\left(\frac{1}{4}+(\phi_i-\phi_j-m)^2\right)\tanh(\pi(\phi_i-\phi_j- m))\nn\\
&&\qquad\qquad+\frac12\left(\frac{1}{4}+(\phi_i-\phi_j+ m)^2\right)\tanh(\pi(\phi_i-\phi_j+ m))\Bigg]\nn\,,\nn\\
\label{eom}
\end{eqnarray}
where $\kappa$ is defined in (\ref{kappadef}).  We have also included a Lagrange multiplier $\mu$ which we set to zero for  $U(N)$, or  adjust so that $\sum _i\phi_i=0$ for  $SU(N)$\,.   Since $m$ and $\lambda^{-1}$ have explicit $r$ dependence,  the theory is superconformal only when these terms are zero.
 
The saddle point equation in (\ref{eom}) is difficult to solve exactly, so we will proceed by considering its weak and strong coupling limits. 
Under some  assumptions the equation simplifies in these limits and can be solved analytically using
standard matrix model techniques (see for example \cite{Marino:2011nm}).
 In order to check the validity of our assumptions, we compare  our  analytical results for the approximate equations with the 
 numerical solutions  of the exact  equations.  
 
 To 
 obtain the numerical solutions we will use an idea similar to  one used in \cite{Herzog:2010hf} for the ABJM matrix model. 
  The $N$ algebraic equations in (\ref{eom}) come from minimizing the free-energy $\cal F$ with respect to the eigenvalues, $-\frac{\partial \cal F}{\partial \phi_i}=0$. Instead of solving this directly,
 we introduce a ``time" dependence for the matrix model eigenvalues $\phi_i (t)$ and solve the ``heat" 
 equation
\begin{equation}
 \tau \frac{d\phi_i}{dt}=-\frac{\partial \cal F}{\partial \phi_i}\,.
 \label{heat}
\end{equation}
At  large time-scales $t\to\infty$, with an appropriate choice of $\tau$ the solution of (\ref{heat}) 
relaxes and approaches  the solution
of the saddle point equations. 

\section{Weak coupling}

In the weak coupling limit ($\la,\tilde \lambda\ll1$) we assume that the separations between eigenvalues are small, 
i.e. $|\phi_i-\phi_j|\ll1$. Under this assumption,
 (\ref{eom}) reduces to
\be\label{weak}
\frac{\pi N}{\tilde\lambda}\left(\phi^2_i+2\kappa\phi_i-\mu\right)&\approx& 2\sum_{j\ne i}\frac{1}{\phi_i-\phi_j}\,.
\ee
In the large-$N$ limit  (\ref{weak}) is well approximated by the integral equation
\be\label{weakint}
\frac{\pi N}{\tilde\lambda}\left(\phi^2+2\kappa\phi-\mu\right)&\approx& 2\pint\frac{\rho(\phi')\,d\phi'}{\phi-\phi'}~,
\ee
where the eigenvalue density is normalized to $\int\rho(\phi)d\phi=1$.

A general solution of (\ref{weak})  has two cuts and we can use standard matrix model technology to find these more
  general solutions.  Defining the resolvent,
\be\label{resolv}
w(\phi)=\int\frac{\rho(\phi')d\phi'}{\phi-\phi'}\,,
\ee
 it is straightforward to show using the equations of motion and  its asymptotic behavior that $w(\phi)$  has the general form
\be\label{weq}
w(\tilde\phi)=\frac{\pi}{2\tilde\lambda}\left(\tilde\phi^2-\kappa^2-\mu-\sqrt{(\tilde\phi^2-\kappa^2-\mu)^2-\frac{4\tilde\lambda}{\pi}\tilde\phi\,{+}B}\,\right)\,,
\ee
where $\tilde\phi=\phi+\kappa$.  It then follows that the eigenvalue density is
\be\label{rhoeq}
\rho(\phi)=\frac{1}{2\tilde\lambda}\sqrt{{-}B+\frac{4\tilde\lambda}{\pi}\tilde\phi-(\tilde\phi^2-\kappa^2-\mu)^2}\,.
\ee
There are four branch points bounding the two eigenvalue cuts, and a free parameter $B$  that adjusts their filling fractions.  

If the Yang-Mills coupling is small in comparison to the CS coupling, then we expect  the relevant solution of (\ref{weakint}) to have a single cut along the real axis.  This corresponds to choosing
\be\label{Beq}
B={-}(\kappa^2+\mu-b^2)(\kappa^2+\mu+3b^2)
\ee
where $b$ satisfies the equation
\be\label{brel}
b(\kappa^2+\mu-b^2)=\frac{\tilde\lambda}{\pi}\,.
\ee
For this choice of $B$ two of the branch points merge and the resolvent  becomes
\be\label{weq1cut}
w(\tilde\phi)=\frac{\pi}{2\tilde\lambda}\left(\tilde\phi^2-\kappa^2-\mu-(\tilde\phi+b)\sqrt{(\tilde\phi-b)^2-2(\kappa^2+\mu-b^2)}\right)\,,
\ee
which gives an eigenvalue density
\be\label{rhoeq1cut}
\rho(\phi)=\frac{1}{2\tilde\lambda}(\tilde\phi+b)\sqrt{\frac{2\tilde\lambda}{b\,\pi}-(\tilde\phi-b)^2}
\ee
between the square-root branch points at $\phi=b-\kappa\pm\sqrt{2(\kappa^2+\mu-b^2)}$.  

In the $SU(N)$ case we have that
\be0=\int d\phi~\phi\rho(\phi)=\frac{\pi}{4\tilde\lambda}(\kappa^2+\mu-b^2)\left((3b-\kappa)(b-\kappa)+\mu\right)\,,
\ee
which leads to $\mu=(\kappa-3b)(b-\kappa)$.  The density then becomes
\be\label{SUNd}
\rho(\phi)=\frac{1}{2\tilde\lambda}(\tilde\phi+b)\sqrt{8b(\kappa-b)-(\tilde\phi-b)^2}\,,
\ee
and the relation in (\ref{brel}) is now
\be\label{brel2}
4b^2(\kappa-b)=\frac{\tilde\lambda}{\pi}\,.
\ee

In both the $U(N)$  and $SU(N)$ cases there is a phase transition when $\lambda$ becomes large enough.  This would occur when, say, the radius $r$ is decreased.  In terms of the densities in (\ref{rhoeq1cut}) and (\ref{SUNd}), this happens when the zero at $\tilde\phi=-b$ coincides with  the left branch point.  In the $U(N)$ case where $\mu=0$, the critical value  occurs when $\kappa^2=3b^2$.  Using (\ref{brel}) this corresponds to 
\be\label{kappaUN}
\kappa=\kappa_{crit}\equiv\sqrt{3}\left(\frac{\tilde\la}{2\pi}\right)^{1/3}\,,
\ee
 from which it follows that
\be
\lambda=\frac{2}{\sqrt{3}}(2\pi)^{7/3}\tilde\lambda^{2/3}\,.
\ee
For $SU(N)$ the critical value happens when $\kappa=\frac{3}{2}b$, and so using (\ref{brel2}) 
\be\label{kappaSUN}
\kappa_{crit}=\frac{3}{2}(\tilde\la/2\pi)^{1/3}\,,
\ee
 and thus
\be
\lambda=\frac{4}{{3}}(2\pi)^{7/3}\tilde\lambda^{2/3}\,.
\ee
If $\kappa>\kappa_{crit}$,  then all eigenvalues are real.  If  $\kappa<\kappa_{crit}$, then s

ome of the eigenvalues are complex.
We will study the phase transitions more closely in section \ref{YMCS}, where we  show that the transition is third order.

\subsection{The weakly coupled pure CS model}
\label{weakcubic}

\begin{figure}
\begin{center}
    \subfigure[Eigenvalues of the $Z_3$ solution for $N=123$, $\tilde\lambda=0.02$]{\label{eigenvalues:Z3}
  \includegraphics[width=50mm,angle=0,scale=1.36]{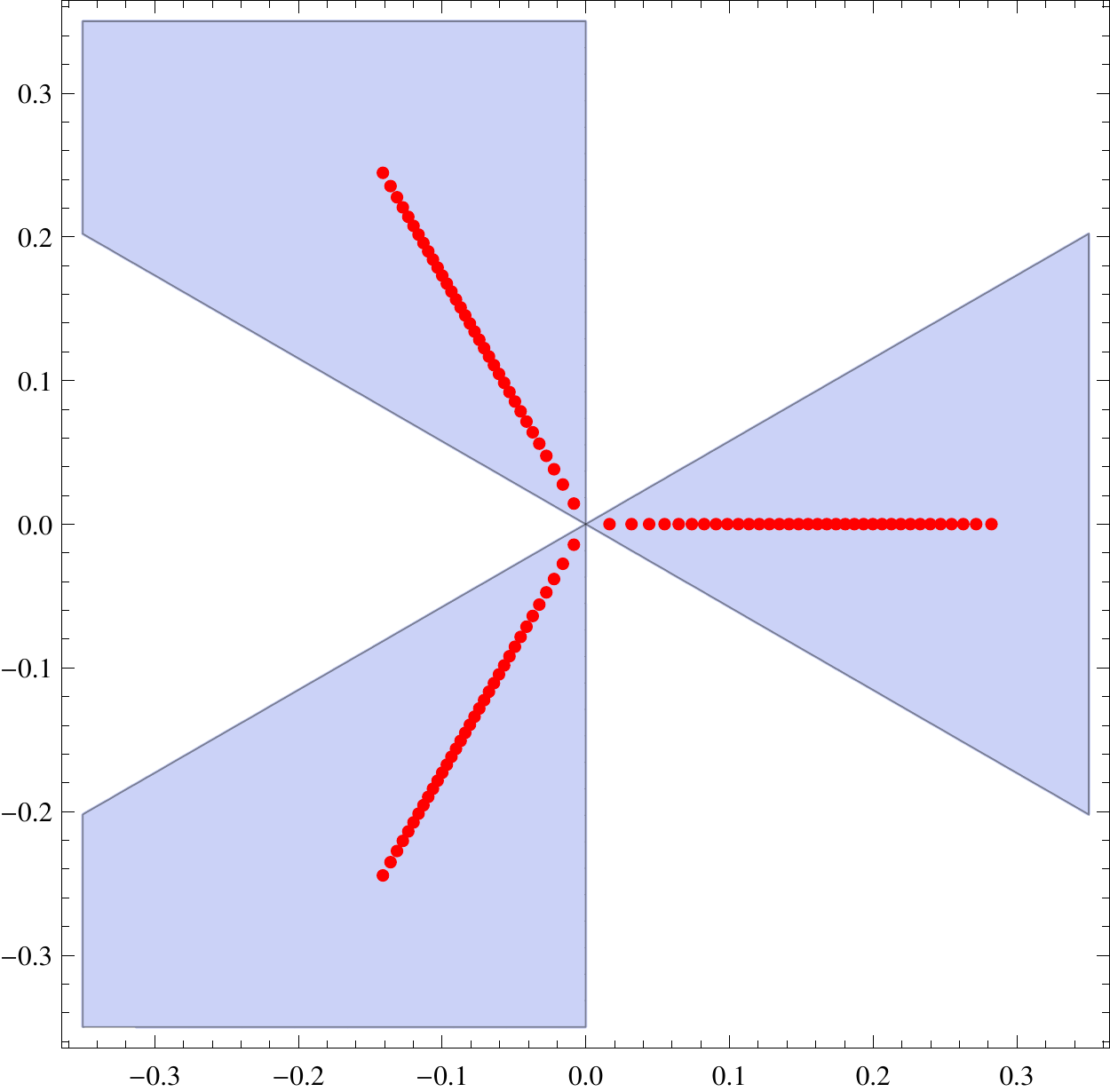}}
  \hspace{7mm}
\subfigure[Eigenvalues of the three single-cut solutions for $N=51$, $\tilde\lambda=0.1$]{\label{eigenvalues:3}
  \includegraphics[width=53mm,angle=0,scale=1.3]{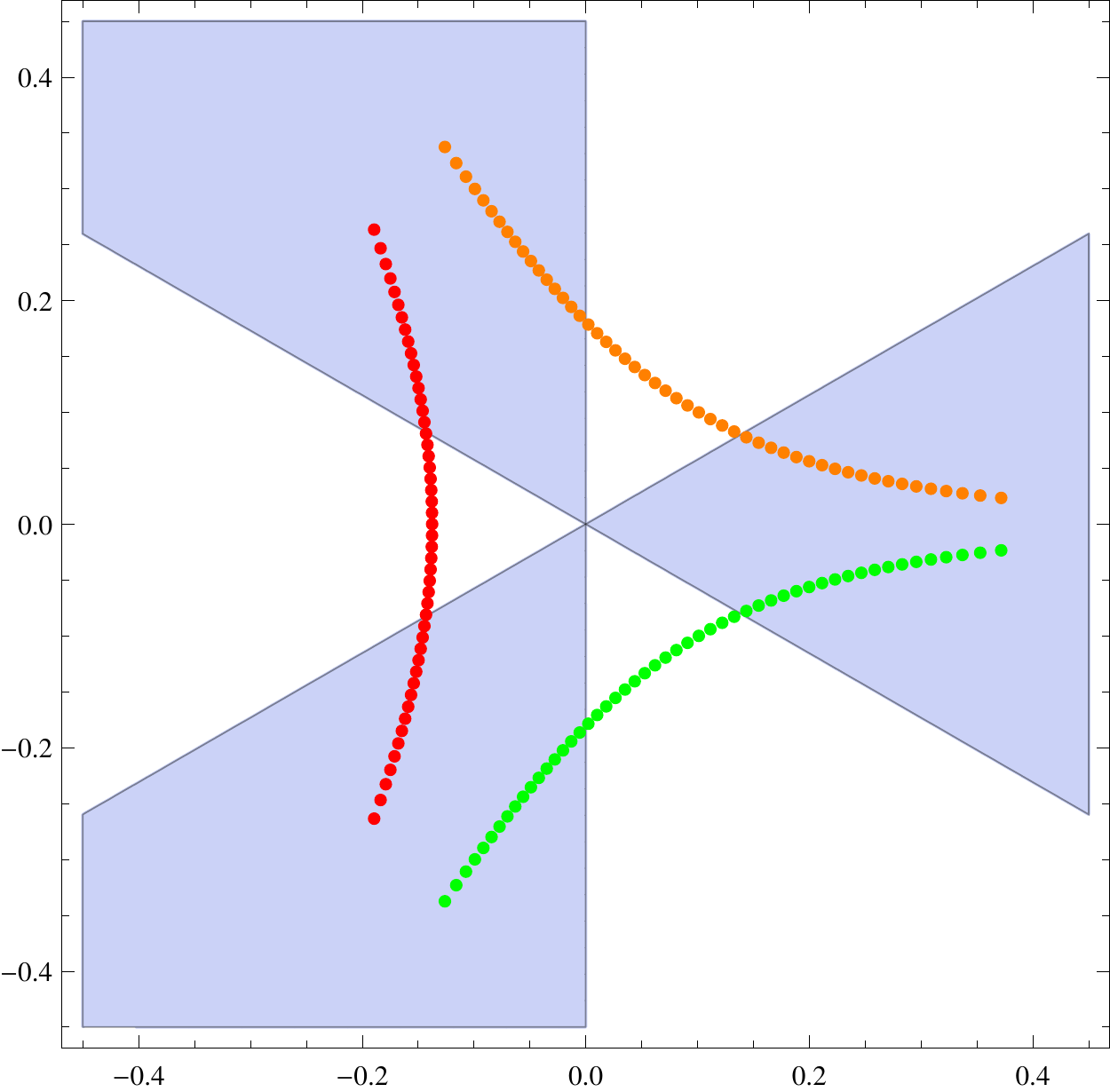}}
\end{center}
\caption{Eigenvalues for the pure CS  model at weak coupling.  The blue regions are 
the integration regions in the complex plane where $\mbox{Re}(\phi^3)>0$ so that the path integral
converges.  The distributions were computed numerically using the heat equation on (\ref{weak}).}
\label{eigenvalues}
\end{figure}

Taking  $\kappa=0$ so that the YM coupling is infinite,  we go beyond the critical point and the matrix model reduces to the pure CS model.    For the weakly coupled $U(N)$ case, the  equations   in (\ref{weak}) and (\ref{weakint}) have an invariance under the $Z_3$ transformation 
$\phi_i\to\omega\phi_i$, $\omega=e^{2\pi i/3}$, hence one can look for solutions that are also $Z_3$-symmetric.
Such solutions will have three 
branches, where the eigenvalues sit at  $\phi_i$,  $\omega\phi_i$ and $\omega^2\phi_i$, with $\phi_i$ positive real. 
We can then write (\ref{weak}) as

\be
\frac{\pi N}{\tilde\lambda}\phi^2_i&\approx& 2\sum_{j\ne i}^{N/3}\frac{1}{\phi_i-\phi_j}+
2\sum_{j}^{N/3}\frac{1}{\phi_i-\phi_j\omega}+2\sum_{j}^{N/3}\frac{1}{\phi_i-\phi_j\omega^2},
\ee
which can be  rewritten as
\be\label{weakZ3}
\frac{\pi N}{\tilde\lambda}\phi^2_i&\approx& 6\sum_{j\ne i}\frac{\phi_i^2}{\phi^3_i-\phi^3_j}+\frac{2}{\,\phi_i}~.
\ee
Letting $\Phi_i=\phi_i^3$, and taking the large
$N$ limit we can turn (\ref{weakZ3}) into the integral equation
\be\label{weakZ32}
\frac{\pi }{\tilde\lambda}&\approx& {2}\pint\frac{\hat\rho(\Phi')}{\Phi-\Phi'}d\Phi'\,,
\ee
where the density of eigenvalues is normalized to  $\int\hat\rho(\Phi)d\Phi=1$.  Using  standard
matrix model techniques, one finds that 
\be
\hat\rho(\Phi)=\frac{1}{2\tilde\lambda}\sqrt{\frac{4\tilde\lambda}{\pi\,\Phi}-1}\,.
\label{rhoZ3}
\ee
In this case the cut runs between the origin and $\Phi=4\tilde\lambda/\pi$.  In terms of $\phi$, there are three cuts   emanating out of the origin and running toward the square root branch points at $\phi=({4\tilde\lambda}/{\pi})^{1/3}\omega^n$ 
for $n=0,1,2$.  The origin is also a square root branch point, with the three directions of the cuts determined by keeping $\rho(\phi)d\phi$  positive definite.  The eigenvalue distribution for this solution is shown in fig.\,\ref{eigenvalues:Z3}.
Because of the $Z_3$ symmetry, the average  of the eigenvalues  is $\langle\phi\rangle=0$, 
thus, there is no
distinction between $U(N)$ and $SU(N)$ for this type of solution.

The $Z_3$ solution can also be obtained from (\ref{weq}) and (\ref{rhoeq}) by setting the filling fraction parameter $B=0$.  In this limit the side of one of the two cuts collides with a branch point of the other cut, leaving three symmetric cuts.

The single-cut solutions obtained from (\ref{weq1cut}) by setting $\kappa=0$ are not $Z_3$-symmetric, but transform into each other under $Z_3$ transformations.  For $U(N)$ the eigenvalue density  has the form in (\ref{rhoeq1cut}) where $b$ satisfies
\be\label{beq}
b^3=-\frac{\tl}{\pi}\,.
\ee
This equation has three roots corresponding to the three different solutions,  as shown in fig.\,\ref{eigenvalues:3}.
If one attempts to generate single-cut $SU(N)$ solutions using the eigenvalue density in (\ref{rhoeq1cut})  with 
\be\label{beqSUN}
b^3=-\frac{\tl}{4\pi}\,,
\ee
 it does not work.  Starting at one of the branch points and following a trajectory such that $\rho(\phi)d\phi$ is positive definite, one finds that the curve runs out to infinity instead of to the other  branch point.  From this we conclude that there are no single-cut solutions for $SU(N)$.

We next consider the free-energy  for the $Z_3$ and the single-cut solutions.  In the large-$N$ limit  the free-energy is given by
\be
F=\frac{kN\pi}{3}\int_{\mathcal{C}} \phi^3\rho(\phi)d\phi-\frac{N^2}{2}\int_{\mathcal{C}} d\phi
d\phi'\rho(\phi)\rho(\phi')\log(\phi-\phi')^2-N^2 C
\label{free:energy:general}
\ee
where the contour $\mathcal{C}$ is determined by the filling parameter $B$ in the resolvent. The last term 
in (\ref{free:energy:general}) comes from  the first subleading term  in the expansion of the full matrix model potential. Carrying out the expansion, one finds
\be
C=\frac{1}{8}\log2 + \log\pi +\frac{7 \zeta(3)}{16\pi^2}
\label{constant:term}
\ee
Details for computing the integrals in (\ref{free:energy:general}) can be be found in Appendix \ref{appendix:weak}, 
where we show that
\be
F&=&N^2\left(\frac{1}{2}-\frac{1}{3}\log\frac{\tilde\lambda}{\pi}-C\right)
\ee
for the $Z_3$  solution and
\be
F&=&N^2\left(\frac{1}{2}-\frac{1}{3}\log\frac{\tilde\lambda}{\pi}+\frac{1}{2}\log2-C\right)
\,.
\label{free:energy:approx}
\ee
for the single-cut solutions. The free-energy for both solutions scales as $N^2$, but the $Z_3$ solution has lower free-energy and is thus the energetically preferable cut configuration.

The cut configuration is actually determined by the choice of an integration contour \cite{Witten:2010cx,Felder:2004uy}.  In order for the path integral to converge  the integration over the eigenvalues must asymptote into the blue regions in 
fig.\,\ref{eigenvalues}.  Each eigenvalue integral  then connects two of the three regions.  One of the regions has to be the one that covers the positive real axis if we are to have a  solution that continuously connects onto the pure Yang-Mills solution.    This would then exclude the single-cut solution shown in red dots in fig.\,\ref{eigenvalues:3}.   If the other end of the integration region is the same for all eigenvalues, then this would correspond to one of the two remaining single-cut solutions.  However, if half the eigenvalues asymptote into the region bordering the positive imaginary axis  and the other half into the region bordering the negative imaginary axis, then this gives the $Z_3$-symmetric configuration\footnote{Note that having the eigenvalues on different contours is  consistent with gauge invariance.}.

\section{Strong coupling with adjoint matter}
 
 We now suppose that the couplings are large, $\la,\tilde\la\gg1$.  We can then assume  that  \mbox{$|\mbox{Re}(\phi_i-\phi_j)|\gg1$} for most $i$ and $j$, in which case we can approximate (\ref{eom}) as
\be
N\frac{ \pi}{\tilde\lambda}(\phi^2_i+2\kappa\phi_i+\mu)=
\left(\frac{9}{4}+ m^2\right)\pi\sum\limits_{j\neq i}\mathrm{sign}(\mbox{Re}(\phi_i-\phi_j))\,.
\label{eom:strong}
\ee
Assuming that the $\mbox{Re}(\phi_i)$ are ordered, we get the relation
\be\label{phieqs}
\phi^2_i+2\kappa\phi_i+\mu=\chi\,\frac{2i-N}{N}\,,
\label{solution:strong}
\ee
and an  eigenvalue density  
\be
\rho(\phi)=\frac{1}{\chi}(\phi+\kappa)\,,
\label{strong:density}
\ee
where
\be\label{chieq}
\chi\equiv{\left(\frac94+m^2\right)\tilde\lambda}\,.
\ee
In the $U(N)$ case with $\mu=0$, this means that the eigenvalues range between
$\phi_-$ and $\phi_+$, where
\be\label{phipm}
\phi_\pm=-\kappa+\sqrt{\kappa^2\pm\chi}\,.
\ee
It is then clear that a transition occurs when the argument of the square root in $\phi_-$ vanishes, which occurs when 
\be\label{kappacrit}
\kappa=\kappa_{crit}\equiv\chi^{1/2}=\frac{\sqrt{9+4m^2}}{2}\,\tilde\la^{1/2}\,.
\ee
  Similar to weak coupling, all eigenvalues lie on the real line when $\kappa$ is above (\ref{kappacrit}), but  some are complex when $\kappa$ is below  (\ref{kappacrit}).

In the $SU(N)$ case we must set
\be\label{rhoint0}
\int_{\phi_-}^{\phi_+}\rho(\phi)\phi\,d\phi=0\,,
\ee
which leads to the relation
\be\label{rhoint0-1}
\sfrac{1}{3}\phi_+^3+\sfrac12\kappa\phi^2_+=\sfrac{1}{3}\phi_-^3+\sfrac12\kappa\phi^2_-\,,
\ee
where $\phi_+$ and $\phi_-$ are again the endpoints of the eigenvalue distribution.  Combining this with the relations
\be\label{phi1phi0}
\phi_+^2+2\kappa\phi_+-\mu&=&\chi\nn\\
\phi_-^2+2\kappa\phi_--\mu&=&-\chi\,,
\ee

we obtain the sytem of three equations that define the endpoints of the cut $\phi_\pm$ and the Lagrange 
multiplier $\mu$.

Introducing new variables $\psi_{\pm}\equiv\phi_+\pm\phi_-$,  we can rewrite (\ref{rhoint0-1}) and (\ref{phi1phi0}) as

\be
\psi_+ \psi_-+2\kappa \psi_-&=&2\chi\,\nn\\
\label{mueq}\psi_+(\psi_++\kappa)+2\mu&=&0\,\nn\\
\label{SUNscp}\frac{4}{3}\,\chi^2+\psi_+(\psi_++2\kappa)^3&=&0\,.
\ee

This last equation has a critical point when
\be
(\psi_++2\kappa)^3+3\psi_+(\psi_++2\kappa)^2=0\,,
\ee
which is satisfied when $\psi_+=-\kappa/2$.  Substituting back into (\ref{SUNscp}), we find 
\be\label{kappasSUN}
\kappa_{crit}=\frac{2\sqrt{2}}{3}\,\chi^{1/2}=\sfrac{1}{3}\sqrt{2(9+4m^2)\tilde\lambda}
\ee
for the critical value.  In section (\ref{YMCS}) we will further study these critical points, where we will show that the phase transition stays third-order for strong coupling.

\subsection{Chern-Simons with adjoint matter}
 
We now assume that $\kappa=0$ such that the theory is pure CS and we are past the phase transition.  As we move to stronger coupling for $\tilde\lambda$, the $Z_3$ symmetry breaks as the determinant factors diverge from the Vandermonde form.  We still expect there to be the analog of the three single-cut solutions in section \ref{weakcubic}, that is one solution with an eigenvalue distribution symmetric about the real axis and the other two  complex conjugates of
each other.  We also expect an analog of the $Z_3$ solution, which has a branch on the real line and two complex branches that are conjugate to each other.  We still assume that  $|\mbox{Re}(\phi_i-\phi_j)|\gg1$  for generic eigenvalues and that the $\mbox{Re}(\phi_i)$ are ordered. Hence, the eigenvalues still satisfy (\ref{solution:strong}) and (\ref{strong:density}), with $\kappa=0$.

For the case of $U(N)$ with $\mu=0$, we see that the the righthand side of (\ref{solution:strong}) is positive if $i\ge N/2$, hence  these eigenvalues are on the positive  real
line and run between the origin and $\phi_+=\sqrt{\tilde\lambda({9}/{4}+m^2)}$. 
However, for $i<N/2$, the righthand side of (\ref{solution:strong}) is negative and the corresponding eigenvalues lie on the  imaginary axis.  If  the imaginary  eigenvalues have the same sign, then the solution connects to one of the complex single-cut solutions at weak coupling and the free-energy is complex.  Alternatively, the eigenvalues could divide up such that the imaginary eigenvalues appear with their conjugates, in which case the free-energy is real.  We will continue to refer to this as the $Z_3$ solution since this is the one that connects to the $Z_3$ weak coupling solution. 

\begin{figure}
\begin{center}
  \includegraphics[width=53mm,angle=0,scale=1.0]{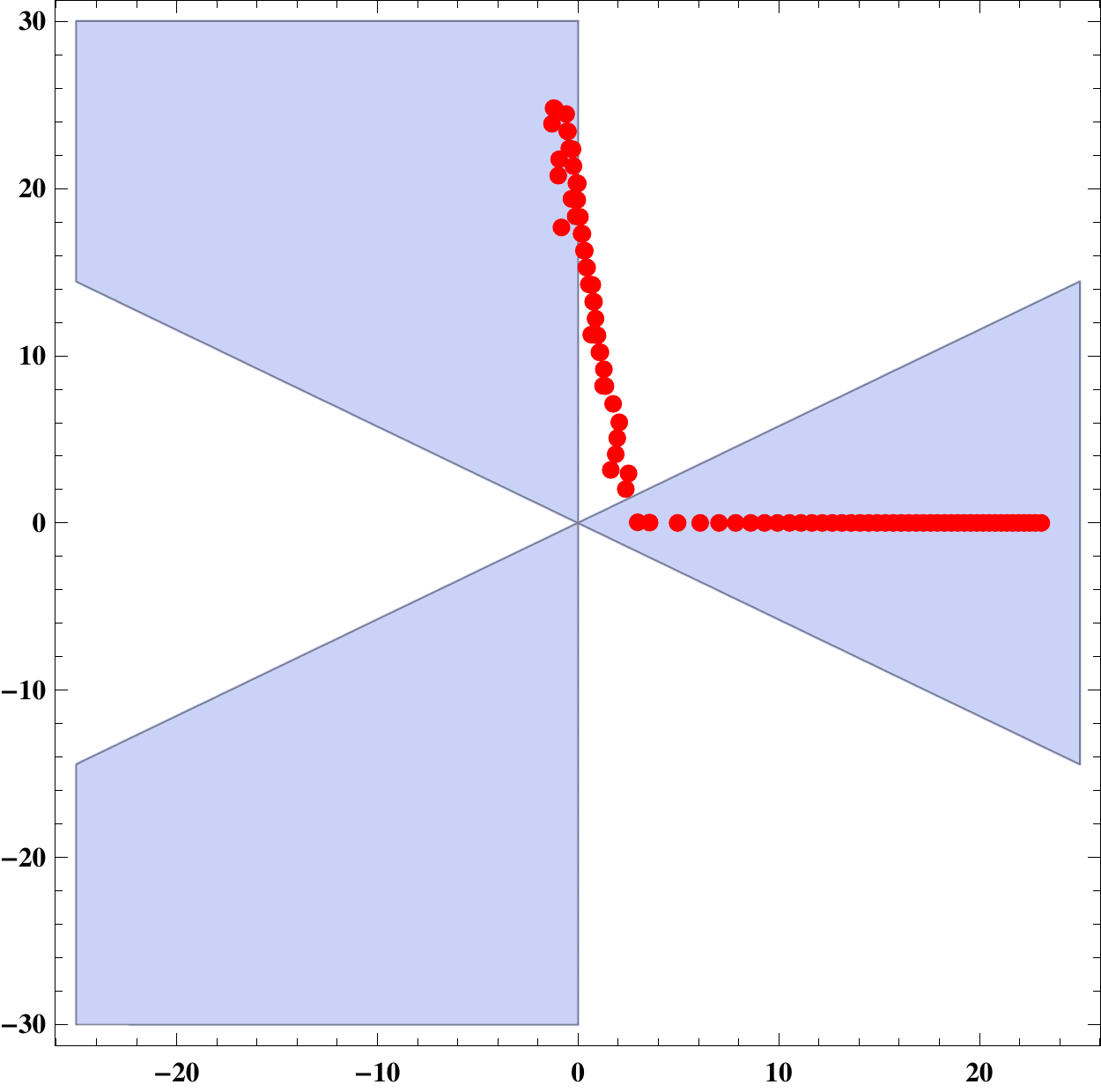}\hspace{25mm}
\end{center}
\caption{Example of strong coupling solution $\tl=750$, $N=87$.}
\label{strong:pic}
\end{figure}

Note that  these branches cannot lie exactly on the 
imaginary axes, since the approximations used for  $\coth\pi(\phi_i-\phi_j)$ and $\tanh\pi(\phi_i-\phi_j)$ 
break down if the real part of the argument is zero.  Instead we should assume that the eigenvalues satisfy 
$|\mbox{Im}(\phi_i-\phi_j)|\gg |\mbox{Re}(\phi_i-\phi_j)|\gg1$, with the ratio of the imaginary to the real 
parts diverging  as $\tilde\lambda\to\infty$.
A numerical solution of (\ref{eom}) at strong coupling  is shown in fig.\,\ref{strong:pic} and confirms these assumptions.
 As shown in the  figure, half of the eigenvalues lie on the  positive real  axis, while the other half spread in the general direction of the  
 positive imaginary axis, but with some separation  along the real axis.  Furthermore, the endpoint toward the imaginary direction is close to the computed value $\phi_-$.

 We also see from the numerical solution that the distribution of the eigenvalues along the imaginary axis is somewhat chaotic.  This is partly due to the poles of the  coth  and tanh functions along the imaginary axis which lead to less numerical precision.  
 
 But more importantly, for high enough $\tilde\lambda$, the single-cut solution no longer exists and the  cut starts bifurcating into multi-cuts along the imaginary direction, with the number of cuts scaling as $\tl^{1/2}$.  We interpret the bifurcating of the cuts as evidence for phase transitions as $\tl^{1/2}$ is increased.
 In appendix\,\ref{appendix:exact} we  show the disappearance of the single-cut solution explicitly  by considering the solution for the special point $m^2=-1/4$, where the matrix model is  solvable analytically.   Here we find a critical value $\tl_c\approx0.976$ where the eigenvalue density goes to zero in the middle of the cut, signifying a splitting into two cuts.    We also argue that a new cut appears when $\sqrt{2}\,\tl^{1/2}$ is increased by $2$.  More generally, we believe that a new cut appears when  $\chi^{1/2}$ increases by $2$.  Each appearance of a new cut signifies a phase transition.  The existence of the phase transitions complicates the search for supergravity duals, and perhaps indicates that they do not exist.

Turning to  $SU(N)$, it is straightforward to show that continuous solutions to (\ref{phieqs}) with $\kappa=0$ do not exist.  Setting $\kappa=0$ in (\ref{rhoint0-1}) and (\ref{phi1phi0}), we are immediately led to
\be
\chi^3+3\mu^2\,\chi=0\,,
\ee
and so
\be
\mu=\pm\frac{i}{\sqrt{3}}\,\chi\,.
\ee
Hence, the endpoints of the eigenvalue distribution consistent with  (\ref{rhoint0-1}) are at
\be
\phi_+=\frac{\sqrt{2}}{3^{1/4}}\,\chi^{1/2}\,e^{\pm i\pi/12}\,,\quad \phi_-=-\frac{\sqrt{2}}{3^{1/4}}\,\chi^{1/2}\,e^{\pm 7i\pi/12}\,.
\ee
But these endpoints cannot be connected by a continuous distribution of eigenvalues because they lie on different branches.  However, we believe there is an approximate solution where
$\mu=\chi$, making  $\phi_i$ imaginary for all $i$.  Assuming that the conjugates also appear, then $\langle\phi\rangle=0$, satisfying the traceless condition.
 
   \begin{figure}
\begin{center}
{
  \includegraphics[width=50mm,angle=0,scale=1.4,natwidth=300,natheight=200]{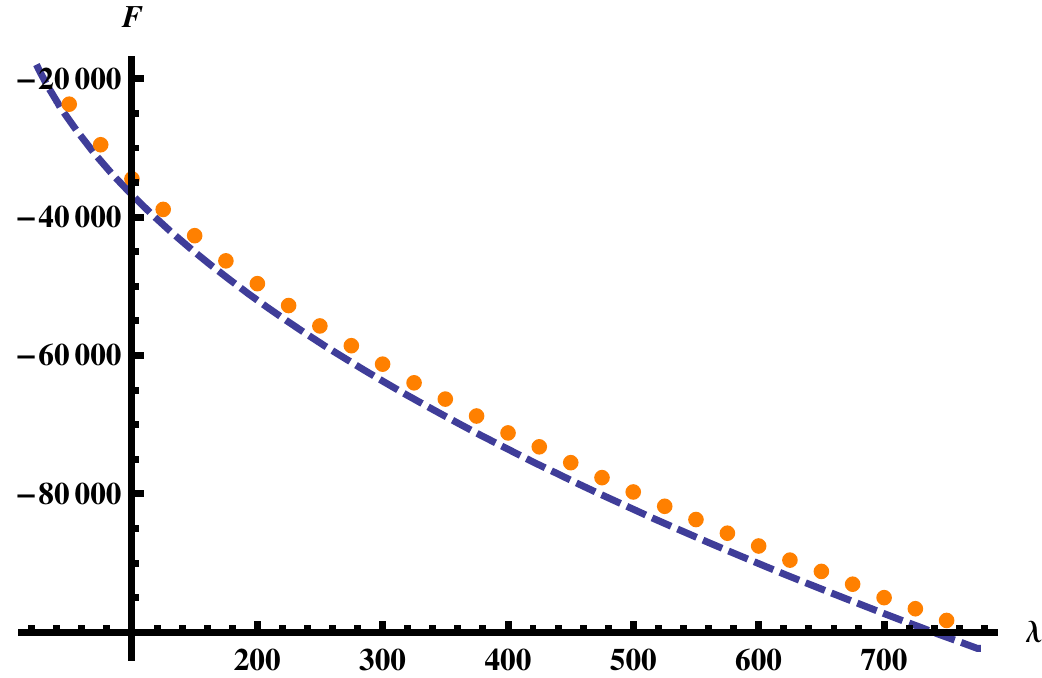}}
\end{center}
\caption{$\tl$-dependence of the free energy  at strong coupling. The orange dots 
represent the numerical solution while the dashed blue line is the solution in (\ref{FE-strongscp}).}
\label{free:energy-strong}
\end{figure}

Returning to the $U(N)$ case and using our assumptions about the eigenvalues,
we can approximate the  free energy  as
\be\label{FE}
 F \equiv -\log Z&\approx& \sum_i\frac{\pi k}{3}\phi_i^3-\frac{(9+4 m^2)\pi}{4}\sum_{i<j}|\mbox{Re}(\phi_i-\phi_j)|\,.
 \ee
 For the $Z_3$ solution this becomes
 \be
 &F\approx&\sum_{i>N/2}\frac{\pi k}{3}\phi_i^3-\frac{(9+4m^2)\pi}{4}\frac{N}{2}\sum_{i>N/2}\phi_i-\frac{(9+4m^2)\pi}{4}\sum_{N/2<i<j}
 (\phi_j-\phi_i)
 \,,
 \label{matrix:free:energy}
\ee
where the leading contributions from the two complex branches have canceled out.  Then plugging 
(\ref{solution:strong}) into (\ref{matrix:free:energy}), we find 
\be\label{FE-strong}
F\approx -\frac{(9+4m^2)^{3/2}\pi}{60}N^2 \tilde\lambda^{1/2}\,,
\ee
which  simplifies to 
\be\label{FE-strongscp}
F\approx -\frac{9\pi}{20}N^2 \tilde\lambda^{1/2}
\ee
 at the superconformal point.
For the solutions that have all of their complex roots either above or below the real line, the free-energy is complex, but the real part matches (\ref{FE-strong}). 

Substituting $\tl=N/k$ into (\ref{FE-strong}) and (\ref{FE-strongscp}), we find  $N^{5/2}$ behavior for the free-energy at strong coupling.  This parallels the strong-coupling behavior of the $5D$ superconformal theories in \cite{Jafferis:2012iv}.   In our case, by adjusting $\kappa$ we can transition between the $N^3$ behavior that one finds for SYM, which is related to the behavior of $6D$ superconformal theories, and $N^{5/2}$ behavior which is expected for $5D$ superconformal theories.  Note that the $SU(N)$ theories will not have the $N^{5/2}$ behavior, because in the leading approximation their eigenvalues lie on the imaginary axis and so their contribution to the free-energy cancels.

As a further check on our results, we  computed the real part of the free-energy numerically as a function of $\tilde\lambda$  for the solution in fig.\,\ref{strong:pic}.   The results of this analysis are shown in  fig.\,\ref{free:energy-strong}.  Here we see that the approximate result in (\ref{FE-strongscp}) accurately reproduces the numerical result.

\section{Wilson Loops}
\label{Wilson}

A supersymmetric  Wilson loop wrapping the equator of $S^5$ can also be obtained from the matrix model in (\ref{vh1loop-intro}). Such loops were considered in  
\cite{Young:2011aa} and \cite{Minahan:2013jwa} for $5D$ $SU(N)$ SYM theory and in \cite{Assel:2012nf} for 
$5D$ superconformal theories. 
One twist to the situation here is that the CS term in the action is odd under charge conjugation,
hence the Wilson loop for the fundamental representation can be different from the Wilson loop in 
the anti-fundamental representation.\footnote{{However in 3d CS theory, which also has an action odd under charge 
conjugation, Wilson loops in the fundamental and anti-fundamental representations behave in the same way 
(see for example \cite{Marino:2011nm}).}}

The  expectation value of the Wilson loop in the fundamental or anti-fundamental representation, after localizing, is the expectation value  in the matrix model (\ref{vh1loop-intro}) \cite{Pestun:2007rz},
\be
\langle W\rangle^{\pm}=\frac{1}{N}\langle \Tr \,e^{\pm2\pi\phi}\rangle\,,
\ee
where the $+$ ($-$) sign refers to the (anti-)\,fundamental representation.
In the large-$N$ limit the back-reaction of this term on the eigenvalue
distribution is negligible, hence the expectation value of the loop is well approximated by
\be
\langle W\rangle^{\pm} = \int d\phi \rho(\phi)e^{\pm2\pi\phi}\,,
\label{wilson:loop:mm}
\ee
where $\rho(\phi)$ is the eigenvalue density computed in the previous sections.

In the rest of this section we consider the Wilson loop for the pure CS models at weak and strong coupling.

 \subsection{Purely cubic model at weak coupling}
 
At weak coupling we have studied two types of solutions, the $Z_3$ solution and the single-cut solutions. Let us consider Wilson loops for these configurations separately.

 \subsubsection{$Z_3$ symmetric solution}
 
The $Z_3$ solution is given by (\ref{rhoeq}) with 
 $B=\kappa=\mu=0$. This solution consists of three branches that can be mapped into each other by  $e^{2\pi i/3}$
 rotations. All three branches contribute to the integral (\ref{wilson:loop:mm}), leading  to the expression
   \be
 \langle W\rangle_{Z_3}^{\pm} =\int\limits_0^{\phi^*}d\phi \frac{\phi^{2}}{2\tilde\lambda}\sqrt{\frac{4\tilde\lambda}
  {\pi \phi^3}-1}\left[e^{\pm2\pi\phi}+e^{\pm\pi\phi(-1+\sqrt{3}i)}+e^{\pm\pi\phi(-1-\sqrt{3}i)}\right]\,,
 \ee
where the first term in the square brackets comes from the integration over the branch on the real line, while the two other terms come from the rotated branches. The endpoint of the cut on the real line sits at $\phi^{*}=({4\tilde\lambda}/{\pi})^{1/3}$.

 The integral results  in the generalized hypergeometric function
  \be
 \langle W\rangle_{Z_3}^{\pm}=~_1 F_3\left(\frac{1}{2}; \frac{1}{3}, \frac{2}{3}, 2; \pm\frac{32\pi^2 \tilde\lambda}{27}\right)\,.
 \ee
 This expression is real and in the limit $\tilde\lambda\ll1$ its log is approximately
  \be
  \log\langle W\rangle^{\pm}_{Z_3}\approx\pm\frac{4\pi^2 \tilde\lambda}{3}
  \label{wilson:asympt:weak:Z3}
 \ee

\subsubsection{Single-cut solutions}

The three single-cut solutions for $U(N)$ have $\kappa=\mu=0$ in (\ref{rhoeq1cut}), with $b$ one of the roots in (\ref{beq}).  The real root corresponds  to the solution symmetric with respect to the
real axis, while the other roots correspond to the single cuts that are completely above or below the real axis.

 For the symmetric solution the Wilson loop is given by the integral
  \be
 \langle W\rangle_{Z_2}^{\pm} = \int \limits_{\phi_1}^{\phi_2} d\phi\frac{i}{2\tilde\lambda}\left(\phi-\left(\frac{\tilde
 \lambda}{\pi}\right)^{1/3}\right)\sqrt{(\phi-\phi_1)(\phi-\phi_2)}~e^{\pm2\pi\phi},
 \label{wilson:weak:integral}
 \ee
 where 
 \be
 \phi_{1}=({\tilde \lambda}/{\pi})^{1/3}(-1- i\sqrt{2})\,,\qquad
\phi_{2}=({\tilde \lambda}/{\pi})^{1/3}(-1+ i\sqrt{2})\,.
\ee
 Defining the new variable  $x=\frac{1}{i\sqrt{2}}\left(({\tilde \lambda}/{\pi})^{1/3}\phi+1\right)$, we can rewrite (\ref{wilson:weak:integral}) as
  \be
 \langle W\rangle_{Z_2}^{\pm}  =-2\int\limits_{-1}^{1}(\sqrt{2} i x-2)\sqrt{1-x^2}e^{a(i \sqrt{2}x-1)}
 \ee
 where $a=\pm(8\pi^2 \tilde\lambda)^{1/3}$. The integral then gives
  \be\label{WZ2}
 \langle W\rangle_{Z_2}^{\pm} =\frac{e^{-a}}{a}\left(a~_0 F_1\left(2,-\frac{a^2}{2}\right)+J_2\left(\sqrt{2}a\right)\right)
 \ee
 where $_0 F_1$ is the confluent hypergeometric function. This expression is real and for $\tl\ll1$ its log  behaves as
 \be
  \log\langle W\rangle_{Z_2}^{\pm}\approx\mp\frac{3}{2} (\pi^2 \tilde\lambda)^{1/3}\,.
  \label{wilson:asymp:weak:Z2}
 \ee

 The Wilson loop  for the other two single-cut solutions can be found 
 by rotating  $\phi$ in (\ref{wilson:weak:integral}) by $e^{\pm2\pi i/3}$. This is equivalent to rotating $a$ in (\ref{WZ2}) by $e^{\pm2\pi i/3}$.   Hence for $\lambda\ll 1$ the log of the Wilson loops for these configurations are
 \be
  \log\langle W\rangle_{n}^{\pm}\approx\mp \frac{3}{2} (\pi^2 \tilde\lambda)^{1/3}e^{2\pi i n/3}\,.
 \label{wilson:asymp:weak:pm}
 \ee
 
 Note that there is significantly different behavior between the Wilson loops for the $Z_3$ solution 
 and the single-cut solutions.  Not only is the power of $\tl$ different between (\ref{wilson:asympt:weak:Z3}) 
 and (\ref{wilson:asymp:weak:Z2}), but so too is the sign.
 
\subsection{Pure CS model at strong coupling}
 \label{section:wils:strong}

 We next consider the  Wilson loops  in the strong coupling limit where $\tilde\lambda\gg 1$.  We first consider configurations where the imaginary eigenvalues are all above the real axis, as in  fig.\,\ref{strong:pic}.
Substituting 
 the  density in (\ref{strong:density}) into the integral (\ref{wilson:loop:mm}) with $\kappa=0$, we obtain
  \be
 \langle W\rangle_{strong}^{\pm}=\int\limits_{C}d \phi~\frac{\phi~ e^{\pm2\pi\phi}}{\chi},
 \ee
 where $\chi$ is defined in (\ref{chieq}) and  the contour $C$ runs along the real axis up to $\phi_+=\chi^{1/2}$ and the imaginary axis to $\phi_-=i\chi^{1/2}$. Evaluating the integral, we obtain a complex result with components

 \be
\mbox{Re}\left( \langle W\rangle_{strong}^{\pm}\right)&=&\frac{1}{2\pi\chi}\left(\left(\pm\chi^{\sfrac12}
-\frac{1}{2\pi}\right)e^{\pm2\pi \chi^{1/2}}+\chi^{\sfrac12}\sin{2\pi\chi^{\sfrac12}}
+\frac{1}{2\pi}\cos{2\pi \chi^{\sfrac12}}\right)\,,\\
\mbox{Im}\left( \langle W\rangle_{strong}^{\pm}\right)&=&\pm\frac{1}{2\pi\chi}\left(\frac{1}{2\pi}\sin{2\pi \chi^{1/2}}
-\chi^{1/2}\cos{2\pi\chi^{1/2}}\right)\,.
\ee

Since $\tilde\lambda\gg1$, there is clearly a significant difference between $\langle W\rangle^+$ and $\langle W\rangle^-$.  In the former case, the real component  is approximately
\be\label{ReW+}
\mbox{Re}\left( \langle W\rangle^+_{strong}\right)\approx\frac{1}{2\pi}\,\chi^{-1/2}\,
e^{2\pi\chi^{1/2}}\,,
\ee
while $\mbox{Im}\left( \langle W\rangle^+_{strong}\right)\rightarrow 0$ at strong coupling when
 $\tl\gg1$.
 Therefore, the log of the Wilson loop is approximately
  \be
 \log\left( \langle W\rangle^+_{strong}\right)\approx 2\pi\,\chi^{1/2}=2\pi \sqrt{\tilde\lambda\left(\frac{9}{4}+m^2\right)}
 \label{asympt:strong}
 \ee
In the case of $\langle W\rangle^-$, the real part does not have the exponentially growing piece in (\ref{ReW+}), therefore its log is much smaller than (\ref{asympt:strong}).

For the $Z_3$  configuration, where each complex eigenvalue appears with its conjugate, $\mbox{Im}\left( \langle W\rangle^\pm\right)$ cancels, while $\mbox{Re}\left( \langle W\rangle^+\right)$   is still dominated by the real end-point, thus the log is also given by (\ref{asympt:strong}).

\section{Chern-Simons quivers}

 All  results  in the previous sections can be generalized to different types of quiver theories.
Some examples of $5D$ quivers where considered in  \cite{Bergman:2012kr}, \cite{Bergman:2013aca}, 
\cite{Kallen:2012zn} and \cite{Minahan:2013jwa}. But these all considered quivers with pure Yang-Mills  terms in the nodes. 
 Here we will consider quivers with $\NN=1$ $U(N)$ Chern-Simons vector multiplets in the nodes 
 and with hypermultiplets in bifundamental representations. These quivers are more similar to the ones 
 considered in \cite{Imamura:2008nn}, \cite{Herzog:2010hf} or in the simplest case of two nodes in ABJM theory
 \cite{Aharony:2008ug}

\begin{figure}
\begin{center}
  \subfigure[Necklace quiver CS theory with equal levels in all nodes]
  {\label{quiver}\includegraphics[width=40mm,angle=0,scale=1.7]{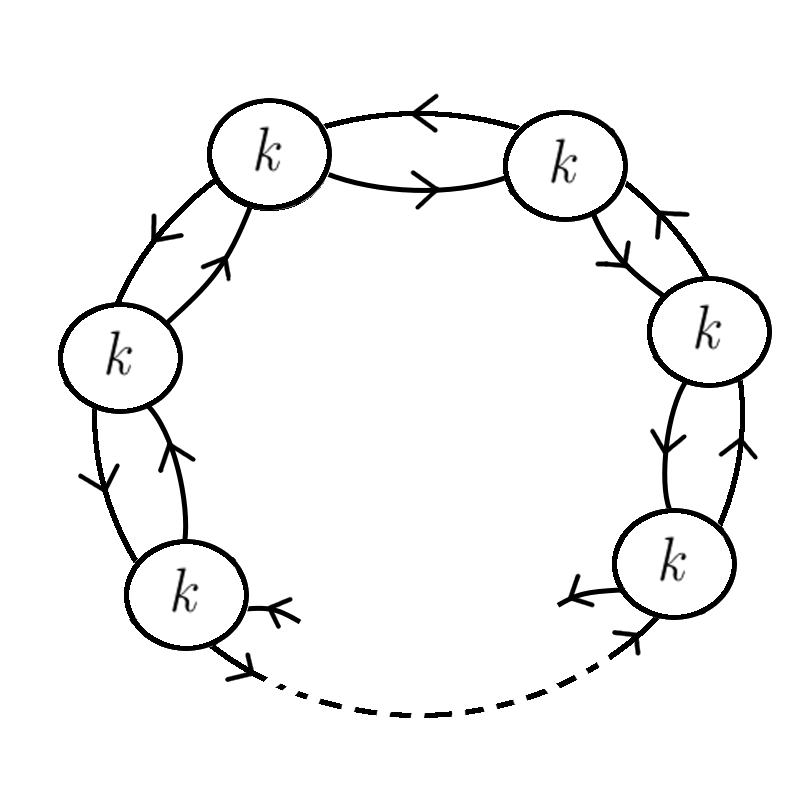}}\hspace{15mm}
  \subfigure[ABJM-like quiver theory]{\label{ABJM}\includegraphics[width=40mm,angle=0,scale=1.0]{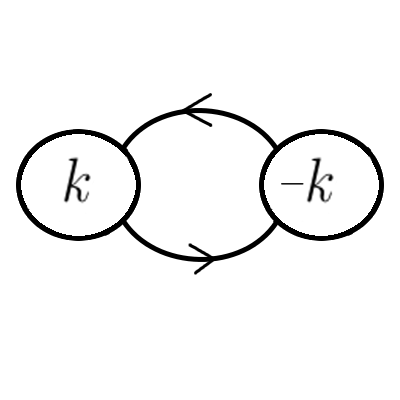}}
\end{center}
\caption{Schematical representations of different Chern-Simons quiver theories}
\end{figure}

The first type of quivers we  consider are necklace quivers with $n$ nodes. Each of the nodes 
 contain $U(N)$ $\NN=1$ Chern-Simons  with equal levels $k$ and matter multiplets in the
 bifundamental representation. This type of quiver is shown schematically  in fig.\,\ref{quiver}.
 The eigenvalues in the saddle-point equations (\ref{eom}) split into 
 $n$ groups $\psi_{i}^{(r)}$ with $r=1,\dots,n$ and $i=1,\dots,N$ and the equations in the planar limit take the form
 \begin{eqnarray}
\frac{\pi N}{\tilde\lambda}(\psi^{(r)}_i)^{2}&=& \pi \Big[\sum\limits_{j\neq i}\left(2- (\psi^{(r)}_i-\psi^{(r)}_j)^2\right)\coth(\pi(\psi^{(r)}_i-\psi^{(r)}_j))\nn
\\
&&\qquad+\Bigg(\sum_j\Big[\sfrac14\left(\sfrac{1}{4}+(\psi^{(r)}_i\ms\psi^{(r\ps1)}_j\ms m)^2\right)\tanh(\pi(\psi^{(r)}_i\ms\psi^{(r\ps1)}_j\ms m))\nonumber\\
&&\qquad\qquad+\sfrac14\left(\sfrac{1}{4}+(\psi^{(r)}_i\ms\psi^{(r\ms1)}_j\ms m)^2\right)\tanh(\pi(\psi^{(r)}_i\ms\psi^{(r\ms1)}_j\ms m))\Big]\Big]\nn\\
&&\qquad\qquad\qquad +(m\to -m)\Bigg)\,.
\label{eom-quiver}
\end{eqnarray}

These equations have the obvious solution $\psi_i^{(r)}=\psi_i^{(s)}$ for all $r$ and $s$. 
The eigenvalues of each quiver satisfy the same saddle-point equations as a single-node theory  (\ref{eom}). 
Thus, the solution at strong coupling is  given by
\be
\phi^{(r)}_{i}=\sqrt{\frac{\left(9+4m^2\right)\tilde\lambda}{4 }\frac{2i-N}{N}}\,,
\label{solution:strong:quiver}
\ee
In the case where $m=0$, we get the same free-energy as in (\ref{FE-strongscp}), multiplied by the number of nodes $n$,
\be\label{FE-strong:quiver}
F\approx -\frac{9\pi}{20}N^2 \tilde\lambda^{1/2}n\,.
\ee
Likewise, for the Wilson loops we get the same asymptotic behavior as for the single-node theory,
 \be
 \log\left( \langle W\rangle_{quiv.}\right)\approx 2\pi \sqrt{\tilde\lambda\left(\frac{9}{4}+m^2\right)}
 \label{asympt:strong:quiver}
 \ee

The solution we have described above is the only one we aware of for the quiver theories.  Furthermore, numerical simulations do not show the
 presence of any other solutions, although a more accurate study of  the equations (\ref{eom-quiver}) could reveal other solutions.

Another type of quiver theory that can be easily generalized from the single-node solution is  an ABJM-like theory with two nodes.
Like ABJM, each node has a $U(N)$ Chern-Simons but with opposite levels $k$ and $-k$. There are also bifundamental  matter fields connecting the two nodes (see fig.\,\ref{ABJM}). 

Denoting the eigenvalues of each node by $\phi_i$ and $\psi_i$, we can write down the equations of 
motion
\begin{eqnarray}
\nonumber
&&\frac{\pi N}{\tilde\lambda}\phi^2_i= \pi \sum\limits_{j\neq i}
\left(2- (\phi_i-\phi_j)^2\right)\coth(\pi(\phi_i-\phi_j))
\\
&&\qquad+\pi \sum\limits_{j}\Bigg[\frac12\left(\frac{1}{4}+(\phi_i-\psi_j-m)^2\right)\tanh(\pi(\phi_i-\psi_j- m))
+(m\to -m)\Bigg]\,,
\label{eom:quiver1}
\end{eqnarray}
\begin{eqnarray}
\nonumber
&&-\frac{\pi N}{\tilde\lambda}\psi^2_i= \pi \sum\limits_{j\neq i}
\left(2- (\psi_i-\psi_j)^2\right)\coth(\pi(\psi_i-\psi_j))
\\
&&\qquad+\pi \sum\limits_{j}\Bigg[\frac12\left(\frac{1}{4}+(\psi_i-\phi_j-m)^2\right)\tanh(\pi(\psi_i-\phi_j- m))
+(m\to -m)\Bigg]\,.
\label{eom:quiver2}
\end{eqnarray} 
 Because of the  symmetry properties of the cubic Chern-Simons term, these equations have the very nice solution $\phi_i=-\psi_i$. Hence,
 the effective equation for the single node takes the form
 \begin{eqnarray}
\nonumber
&&\frac{\pi N}{\tilde\lambda}\phi^2_i= \pi \sum\limits_{j\neq i}
\left(2- (\phi_i-\phi_j)^2\right)\coth(\pi(\phi_i-\phi_j))
\\
&&\qquad+\pi \sum\limits_{j}\Bigg[\frac12\left(\frac{1}{4}+(\phi_i+\phi_j-m)^2\right)\tanh(\pi(\phi_i+\phi_j- m))
+(m\to -m)\Bigg]\,.\nn\\
\label{eom:quiver:final}
\end{eqnarray}
Notice that here we cannot  assume that \mbox{$|\mbox{Re}(\phi_i-\phi_j)|\gg1$} for generic eigenvalues since the quadratic terms $(\phi_i-\phi_j)^2$ and $(\phi_i+\phi_j)^2$ on the r.h.s
do not cancel each other. This makes it difficult to find  approximate equations of motion with analytic solutions.

\begin{figure}
\begin{center}
    \subfigure[First kind of solutions to (\ref{eom:quiver:final})  $N=50$, $\tilde\lambda=1000$, $m=0$]{\label{eigenvalues:ABJM1}
  \includegraphics[width=50mm,angle=0,scale=1.13,natwidth=400,natheight=400]{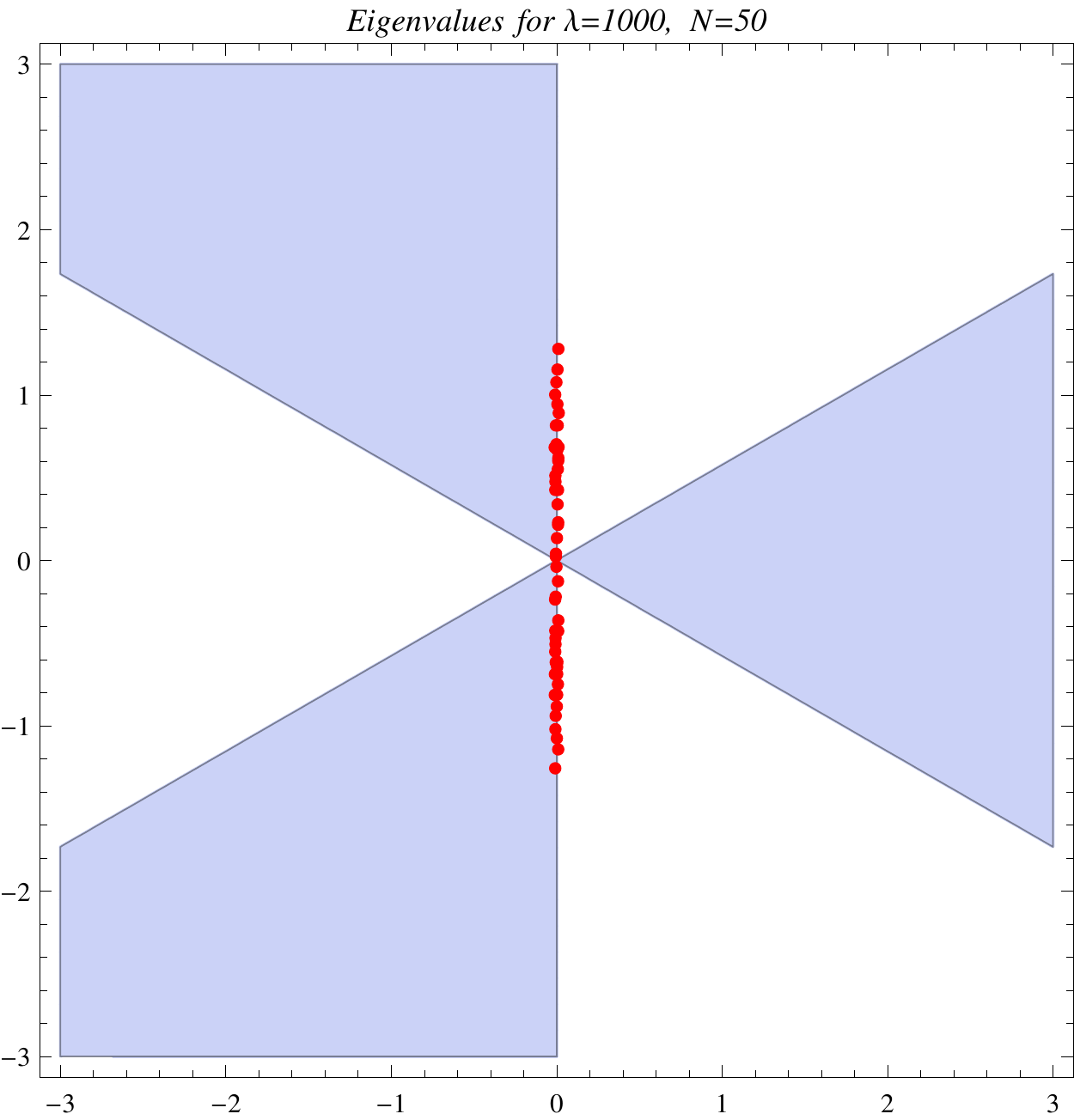}}
  \hspace{7mm}
\subfigure[First kind of solutions to (\ref{eom:quiver:final})  $N=51$, $\tilde\lambda=1000$, $m=0$]{\label{eigenvalues:ABJM2}
  \includegraphics[width=53mm,angle=0,scale=1.09,natwidth=400,natheight=400]{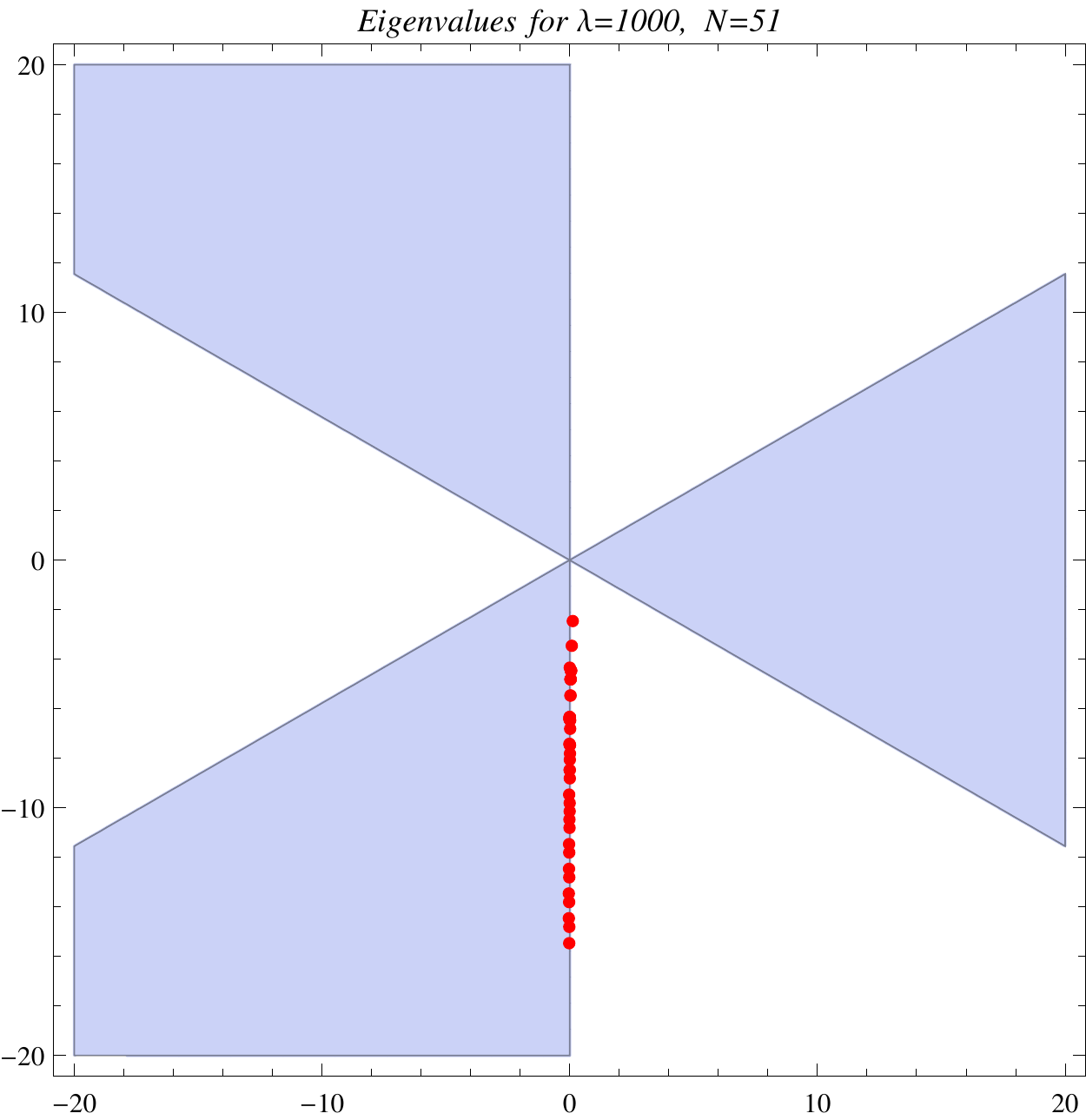}}
\end{center}
\caption{Eigenvalues for the ABJM-like quiver model at strong coupling.  The blue regions are the integration regions in the complex plane where $\mbox{Re}(\phi^3)>0$ so that the path integral converges.}
\label{eigenvalues:ABJM}
\end{figure}

Though we can't say  much about the behavior of the solutions of (\ref{eom:quiver:final}) from 
analytical calculations, we were able to find different numerical solutions of this quiver model. The 
results of the numerical simulations are shown in   fig.\,\ref{eigenvalues:ABJM}. For these solutions we can clearly see that the eigenvalues do not satisfy \mbox{$|\mbox{Re}(\phi_i-\phi_j)|\gg1$}.

For the solution in fig.\,\ref{eigenvalues:ABJM1} the eigenvalues are distributed symmetrically about the real axis and lie close to, but not exactly on the imaginary axis.   Furthermore, the separation between the endpoints stays finite even for large  $\tilde\lambda$, a behavior that is also seen for
 the solution in (\ref{rhophi:symm}) for  pure Chern-Simons  with $m^2=-1/4$. For the  solution shown in  fig.\,\ref{eigenvalues:ABJM2} 
the eigenvalues also lie close to imaginary axis. However  they 
are clearly not symmetric with respect to real axis as they all lie in the lower half-plane. In this case the distance 
between the endpoints increases with  increasing $\tilde\lambda$. 

Since all the eigenvalues for these solutions lie close to the  imaginary axis,  the real part of the free energy cannot go beyond the usual $N^2$ dependence.

 \section{The SYM-CS phase transition}
\label{YMCS}

In this section we elaborate on the phase transition between super Yang-Mills and Chern-Simons behavior.  The main result  is that the phase transition is third order for both weak and strong coupling. 

\subsection{$U(N)$}
We start with the $U(N)$ theory at weak coupling.  Part of this analysis has previously appeared  in the context
of triangulated surfaces in 2D gravity \cite{David:1984tx,Ambjorn:1985az,Kazakov:1985ea,Boulatov:1986jd}, 
but we include it for completeness.  We wish to explore the behavior of the free-energy  near the critical point,  
$\kappa^{2}=3\,b^2$, where $\kappa$ and $b$ satisfy  (\ref{brel}) with $\mu=0$.  We first write $\kappa$ and $B$ in terms of two small parameters $\eps$ and $\delta$,
\be\label{kBeq}
\kappa^2={3}\,L^{2/3}-\eps\,,\qquad B=-12L^{4/3}+4 L^{2/3}\eps+ L^{1/3}\delta\,,
\ee
where $L\equiv (\tl/2\pi)=\kappa_{crit}^3/(3\sqrt{3})\,$.  
The free-energy in the large $N$ limit is given by \footnote{In this section we shift the free energy by  $-\sfrac23\kappa^3$ from (\ref{free:energy:general}).  This will have no effect on the singular terms of the free energy expansion,  but will make some of the expressions nicer.}
\be
F=N^2\left(\frac{1}{2L}\int_{\mathcal{C}} d\phi\rho(\phi)(\sfrac13 \phi^3+\kappa\phi^2-\sfrac23\kappa^3)-\frac{1}{2}\int_{\mathcal{C}} d\phi
d\phi'\rho(\phi)\rho(\phi')\log(\phi-\phi')^2\right)\,,
\label{free:energy:weak}
\ee
where we subtracted off a constant piece to simplify expressions, but will otherwise not effect the phase structure.
Using the more general expression for $\rho(\phi)$ in (\ref{rhoeq}), the free-energy, as an expansion in $\eps$ and $\delta$, is found  to be
\be\label{FEphase}
F=N^2\left(-\frac{3}{4}-\frac{1}{3}\log L+\frac{3}{4}\,L^{-2/3}\eps-\frac{1}{8}\,L^{-4/3}\eps^2-\frac{1}{40}\,L^{-5/3}\eps\,\delta+\dots\right)\,,
\ee
where the expression is valid on either side of the phase transition.

Below the transition we have $\eps<0$ with a single-cut eigenvalue distribution.  Hence, $B$ has the form in (\ref{Beq}).  If we write $b=L^{1/3}+\beta$, then (\ref{Beq}) implies
\be\label{Bexp}
B=-12L^{4/3}+8L^{2/3}\eps+12 L^{2/3}\beta^2+4L^{1/3}\eps\,\beta+12L^{1/3}\beta^3+\dots\,,
\ee
while (\ref{brel}) reduces to
\be
L^{1/3}\eps+3L^{1/3}\beta^2+\eps\,\beta+\beta^3=0\,.
\ee
Solving this last equation for $\beta$ in terms of $\eps$ and substituting into (\ref{Bexp}), we find
\be
B=-12L^{4/3}+4L^{2/3}\eps+8L^{1/3}(-\eps/3)^{3/2}+\dots\,.
\ee
Comparing this equation with (\ref{kBeq}), we find that $\delta\approx 8(-\eps/3)^{3/2}$, and so the free-energy becomes
\be\label{FEbelow}
F=N^2\left(\mbox{regular terms}+\frac{1}{15\sqrt{3}}L^{-5/3}\left(-{\eps}\right)^{5/2}+\dots\right)\,.
\ee
Hence, because  the third derivative of $F$ diverges at this point, there is a third-order phase transition at $\eps=0$.

Let us now continue above the transition to $\eps>0$.    
We assume that the eigenvalues lie on the symmetric two-cut solution 
that connects to the $Z_3$ solution as $\kappa\to0$.  At the critical point, 
three of the four branch points in (\ref{rhoeq}) meet at ${\tilde\phi}=-L^{1/3}$. 
As we move away from the critical point by increasing $\eps$, the branch points spread apart,
and the density near these points is approximately
\be
\rho(\delta\phi)\approx\frac{1}{4\pi L^{5/6}}\sqrt{4\delta\phi^3+4\eps\delta\phi-\delta}\,,
\ee
where $\delta\phi={\tilde\phi}+L^{1/3}$.  We can  shift one of the branch points to zero by setting $\delta\phi=\Delta\phi+x$, where $\Delta\phi$ satisfies the equation 
\be\label{Dphieq}
4\Delta\phi^3+4\eps \Delta\phi-\delta=0\,.
\ee
In terms of $x$, the density is
\be
\rho(x)=\frac{1}{2\pi L^{5/6}}\sqrt{x(x^2+3\Delta\phi x+3\Delta\phi^2+\eps)}\,.
\ee
Assuming that $\eps>0$, we see that two of the branch points are at
\be\label{xend}
x=-\frac{3\Delta\phi}{2}\pm\frac{i\Delta\phi}{2}\sqrt{3+\frac{4\eps}{\Delta\phi^2}}\equiv r e^{\pm i\theta}\,.
\ee
In the limit that $\eps\to0$, $\Delta\phi$ , $\delta$ and $r$ all approach zero. 

To determine the correct value of $\theta$, we now insist that the integral of $\rho(x)$ from $x=0$ to $x=re^{i\theta}$ is positive definite.
We can do the integral, which gives
\be\label{rhoxint}
&&\int_{0}^{re^{i\theta}}\rho(x)dx=\frac{1}{2\pi L^{5/6}}\int_{0}^{re^{i\theta}}dx\sqrt{x(x^2-2r{x}\cos\theta+r^2)}\nn\\
&=&\frac{r^{5/2}}{15\pi L^{5/6}}\sqrt{-2i\sin \theta}\left(2(2\cos{2}\,\theta-1)E\left(\frac{1}{1-e^{2i\theta}}\right)+\left(1-2e^{-2i\theta}\right) K\left(\frac{1}{1-e^{2i\theta}}\right)\right)\,,\nn\\
\ee
where $K$ and $E$ are the complete elliptic integrals of the first and second kind.  We  then adjust $\theta$ such that (\ref{rhoxint}) is positive real.  This can be done numerically, where we find
\be
\theta\approx (0.637775)\,\pi\,.
\ee
Hence the endpoints lie in the second and third quadrants. 

It then follows from (\ref{xend}) that 
\be
\Delta\phi^2=\frac{{4}\,\eps}{9\tan^2\theta-3}\,,
\ee
which then can be used in (\ref{Dphieq}) to give
\be
\delta=\frac{8\eps^{3/2}}{(9\tan^2\theta-3)^{3/2}}(9\tan^2\theta+1)\approx(1.40907)\,\eps^{3/2}\,.
\ee
Substituting this into (\ref{FEphase}), the free-energy above the transition is given by
\be\label{FEabove}
F=N^2\left(\mbox{regular terms}-(0.035223)\,L^{-5/3}\,\eps^{5/2}+\dots\right)\,.
\ee
Curiously, the coefficient  of the $\eps^{5/2}$ term in (\ref{FEabove}) is within 10\% of the coefficient of the $(-\eps)^{5/2}$ term in (\ref{FEbelow}).  

Because of the sign in front of  the singular term,  the free-energy in (\ref{FEabove}) is lower than the real part 
of the free-energy of the one-cut solution.  This latter case is found by analytically continuing $\eps$ 
in (\ref{FEbelow}) to the positive real axis.  Hence the singular term is imaginary.  The regular terms are the same in (\ref{FEbelow}) and (\ref{FEabove}), showing that the two-cut solution is energetically favorable.

Turning now to the phase transition at strong coupling,  we have that the density and endpoints of the integration are given by (\ref{strong:density}) and (\ref{phipm}) respectively.  If we are in the YM phase with $\kappa>\kappa_{crit}$, where  $\kappa_{crit}$ is defined in (\ref{kappacrit}), then
the free-energy is  well approximated by
\be\label{FEint}
F\approx N^2\left(\frac{\pi}{\tl}\int_{\phi_-}^{\phi_+}d\phi\left(\sfrac13\phi^3+\kappa\phi^2-\sfrac23\kappa^3\right)\rho(\phi)-\frac{(9+4m^2)\pi}{8}\int_{\phi_-}^{\phi_+} d\phi d\phi'|\phi-\phi'|\rho(\phi)\rho(\phi')\right)\,.\nn\\
\ee
Using the density and endpoints in  (\ref{strong:density}) and (\ref{phipm}), we find
\be\label{FEsUN}
F&\approx& N^2\frac{8\pi}{15(9+4m^2)\,\tl^2}\left(\left(\kappa^2-\frac{(9+4m^2)\tl}{4}\right)^{5/2}-\left(\kappa^2+\frac{(9+4m^2)\tl}{4}\right)^{5/2}\right)\nn\\
&=&N^2\left(\mbox{regular terms}+\frac{8\pi}{15(9+4m^2)\,\tl^2}\,(-\eps)^{5/2}\right)\,,
\ee
where 
\be\label{epseq}
\eps=\kappa^2_{crit}-\kappa^2\,,
\ee
{and where $\kappa^2_{crit}=(9+4 m^2)\tilde\lambda/4$.}  Hence, the third-order phase transition persists at strong coupling.

On the CS side of the transition, the $\phi_-$ integration boundary in (\ref{FEint}) should be replaced by $0$.  In this case the free-energy is only made up of regular terms.

\subsection{$SU(N)$}
The $SU(N)$ theory has a Lagrange multiplier that could potentially change the nature of the phase transition.  Here we show that although the details differ from the $U(N)$ case, the phase transition stays third order.

 For weak coupling, we can carry out a similar analysis as for (\ref{FEbelow}) in the $U(N)$ case, but also including
 the Lagrange multiplier $\mu$.   We will only consider the system in the YM phase, in which case we can invoke the 
 single-cut density in  (\ref{rhoeq1cut}).  Substituting this into (\ref{free:energy:weak}), we can write $F$ as
\be\label{F1cut}
F=N^2\left(-\frac{1}{2}\log\frac{L}{b}+\frac{L}{12 b^3}+\frac{L^2}{96 b^6}-\frac{\kappa^3}{3L}+\frac{12 b^2 \kappa-3 b\kappa^2}
{24 b^3}+\frac{3}{8}\right)
\ee
where $b$ satisfies (\ref{brel2}).  The critical value for $\kappa$ is given in (\ref{kappaSUN}), which in terms of $L$ is
$\kappa_{crit}=\sfrac32 L^{1/3}$.
Using a slightly different parameterization than we did for the $U(N)$ case, we set $\kappa=\kappa_{crit}-\eps'$.  After substituting this into (\ref{brel2}), we can write the series expansion for $b$ near the critical point as
\be
b&=&L^{1/3}+\sqrt{\frac{2}{3}}L^{1/6}(-\eps')^{1/2}+\frac{4}{9}(-\eps')+\frac{5\sqrt{2}}{27\sqrt{3}}L^{-1/6}(-\eps')^{3/2}\nn\\
&&\qquad\qquad\qquad\qquad+\frac{4}{243}L^{-1/3}(-\eps')^2-\frac{7}{243\sqrt{6}}L^{-1/2}(-\eps')^{5/2}+\dots\,.
\ee
Putting this expression for $b$ into (\ref{F1cut}) we find
\be
F=N^2\left(\mbox{regular terms}+\frac{4\sqrt{2}}{15\sqrt{3}}L^{-5/6}(-\eps')^{5/2}\right)\,,
\ee
hence, the phase-transition is third order in the weak-coupling limit.

We can analyze the behavior at strong coupling by including the Lagrange multiplier $\mu$ in the eigenvalue density that appears in (\ref{FEint}).  This modifies the  the first line of (\ref{FEsUN}) to  to
\be\label{FEsSUN}
 F&\approx& N^2\frac{4\pi}{3(9+4m^2)\,\tl^2}\Biggl(\frac{2}{5}\Biggl[\biggl(\kappa^2+\mu-\frac{(9+4m^2)\tl}{4}\biggr)^{5/2}-\biggl(\kappa^2+\mu+\frac{(9+4m^2)\tl}{4}\biggr)^{5/2}\Biggr]\nn\\
 &&\qquad\qquad-\mu\Biggl[\biggl(\kappa^2+\mu-\frac{(9+4m^2)\tl}{4}\biggr)^{3/2}-\biggl(\kappa^2+\mu+\frac{(9+4m^2)\tl}{4}\biggr)^{3/2}\Biggr]\Biggr)
\ee
Again writing $\kappa=\kappa_{crit}-\eps'$, where $\kappa_{crit}$ is given in (\ref{kappasSUN}), and using (\ref{mueq}) and (\ref{SUNscp}) we can expand $\mu$ near the critical point,
\be
\mu=\frac{1}{8}\kappa_{crit}^2-\frac{1}{2}\kc(-\eps')+\frac{\sqrt{2}}{\sqrt{3}}\kc^{1/2}(-\eps')^{3/2}-\frac{7}{12}(-\eps')^2+\frac{7}{18\sqrt{6}}\kc^{-1/2}(-\eps')^{5/2}+\dots\nn\\
\ee
Inserting this into (\ref{FEsSUN}) and expanding, we find
\be
F\approx N^2\left(\mbox{regular terms}+\frac{4\pi}{5\sqrt{6}}\,\kc^{1/2}\,\tl^{-1}(-\eps')^{5/2}\right)\,,
\ee
hence, the transition stays third order at strong coupling.

\subsection{Wilson loops at the phase transition}

Wilson loops are useful for investigating phase transitions in gauge theories.  In this section we explore how the phase transition affects the Wilson loop at strong coupling.  As we argued in section \ref{Wilson}, the behavior of Wilson loops in the fundamental representation can differ from those in the antifundamental representation.

For a $U(N)$ gauge theory at strong coupling, the two types of Wilson loops are given by
\be
\langle W\rangle^{\pm}=\int_{\phi_-}^{\phi_+} d\phi\,\rho(\phi)e^{\pm2\pi\phi}\,,
\ee
where $\rho(\phi)$ is given by (\ref{strong:density}) and $\phi_{\pm}$ by (\ref{phipm}).  The integral is easily done, resulting in
\be
\langle W\rangle^{\pm}=\pm\frac{2}{\pi(9+4m^2)\tl}\left(\left(\phi_++\kappa\mp\frac{1}{2\pi}\right)e^{\pm2\pi\phi_+}-\left(\phi_-+\kappa\mp\frac{1}{2\pi}\right)e^{\pm2\pi\phi_-}\right)\,.
\ee
If we are just below the transition, then to leading order this becomes
\be
\langle W\rangle^{\pm}\approx
\pm\frac{1}{2\pi\kc^2}\left(\left(\sqrt{2}\kc\mp\frac{1}{2\pi}\right)e^{\pm2\pi(\sqrt{2}-1)\kc}-\left((-\eps)^{1/2}\mp\frac{1}{2\pi}\right)e^{\pm2\pi(-\kc+(-\eps)^{1/2})}\right)\nn\\
\ee
where $\eps$ is defined in (\ref{epseq}).

Taking the log, we get
\be
\log(\langle W\rangle^+)\approx 2\pi(\sqrt{2}-1)\kc-\frac{(2\pi)^2}{3\sqrt{2}\kc}e^{-2\pi\sqrt{2}\kc}(-\eps)^{3/2}+\mbox{regular terms}
\ee
Here we see that the singular term is exponentially suppressed at large coupling.  If instead we consider the other Wilson loop, we find
\be
\log(\langle W\rangle^-)\approx 2\pi\kc+\frac{(2\pi)^3}{3}(-\eps)^{3/2}+\mbox{regular terms}\,.
\ee
Hence this loop is much more sensitive to the transition.

\section{Discussion}

In this paper we have studied the matrix model obtained from $5D$ supersymmetric 
SYM-CS theory on $S^5$. We  solved the model  in both the  weak and 
strong coupling limits. We found for an appropriate choice of contour that the free-energy of the $U(N)$ pure CS theory has the behavior
\be
F  ~~\sim &   - N^2 \log{\tilde\lambda}, ~~ \tilde\lambda\ll 1\\
        &   - N^2 \sqrt{\tilde\lambda}, ~~~ \tilde\lambda\gg 1
\ee
The $U(N)$ CS theory is a superconformal fixed point and the  $N^{5/2}$ behavior at strong coupling is similar to the fixed points in the $USp(N)$ models studied in \cite{Jafferis:2012iv}.

However, we have also argued that there exists a series of phase transitions for increasing $\tl$, making the existence of a supergravity dual, at the very least, problematic.
Accumulating phase transitions have also appeared in $4D$ $\mathcal{N}=2^*$ theories \cite{Russo:2013qaa,Russo:2013kea}, $3D$ massive Chern-Simons theories \cite{Barranco:2014tla,Russo:2014bda}, and mass-deformed ABJM theories \cite{Anderson:2014hxa}.  Unlike the $5D$ pure CS model, these theories are not superconformal, still, there might be interesting connections between the different matrix models that one can explore.

We have also shown the existence of a third order phase transition between an SYM phase and a CS phase when the SYM 
coupling reaches a critical value.  The phase transition exists for any positive $\tl$, and for both 
$U(N)$ and $SU(N)$.   At weak coupling the matrix model and the phase transition are precisely what one finds 
for triangulations of surfaces in $2D$ gravity.  One important feature of the $2D$ gravity studies is the presence 
of a double scaling limit \cite{David:1988hj,Brezin:1990rb,Douglas:1989ve,Gross:1989vs},  which should also exist 
for large $\tl$ where the relation to random surfaces is less obvious.  It would be interesting to discover a more 
concrete connection between the double scaling limit and the  $5D$ SYM-CS theory or even six-dimensional 
superconformal theories.

\section*{Acknowledgements}

We thank Maxim Zabzine for many discussions and early collaboration on this work.
This research is supported in part by
Vetenskapsr{\aa}det under grant \#2012-3269.
JAM thanks the CTP at MIT  for kind
hospitality during the course of this work.

\appendix 
\section{Numerical analysis details}
\label{appendix:numerics}

We use the heat-like equation (\ref{heat}) to obtain numerical solutions of the exact equations 
of motion (\ref{eom}).  But the weak coupling limit in   (\ref{weak}) possesses a $Z_3$-symmetry in the complex $\phi$ plane, while the heat equation breaks the symmetry.  This 
complicates numerical simulations  when there are multiple solutions.  In this appendix we briefly
describe how we  modify (\ref{heat}) in order to obtain the different type of solutions.

\subsection{{Single-cut solution}}

There are three different linearly independent single-cut solutions of the form 
 (\ref{weq}) which are related by $2\pi/3$ rotations in the complex $\phi$-plane. These  solutions are shown with different colors in fig.\,\ref{eigenvalues:3}.

 To obtain the different solutions we need to tune 
 $\tau_1$ so that the heat equation naturally drives the eigenvalues to a particular solution.  For example, the heat equation will evolve toward the symmetric one-cut solution  if we choose $\tau$ to be  positive real.
 After obtaining one solution we can get  the others  using 
 $\tau_2=\omega\tau_1$ and $\tau_3=\omega^2\tau_1$, where $\omega=e^{2\pi i/3}$.

\subsection{{$Z_3$ solution}}
 
  In order to obtain the $Z_3$ solution (\ref{rhoZ3})
 we divide the eigenvalues into three equal groups, and use 
 a different $\tau$ in the heat equation (\ref{heat}) for each group of eigenvalues. Then our equations  look like
\begin{equation}
 \tau_i \frac{d\phi_i}{dt}=-\frac{\partial \cal F}{\partial \phi_i}\,.
\end{equation}
with 
\begin{eqnarray}
 \tau_i & = & \tau , ~~~~ -\frac{N}{2}\leq i\leq -\frac{N+4}{6};\nn\\
        & = & \omega\tau, ~~~~  -\frac{N-2}{6}\leq i \leq \frac{N-2}{6} ;\nn\\
        & = & \omega^2\tau, ~~~~  \frac{N+4}{6}\leq i \leq \frac{N}{2} ;\nn\\
\end{eqnarray}
This trick preserves the $Z_3$-symmetry of the algebraic 
equation $-\frac{\partial \cal F}{\partial \phi_i}=0$ inside the heat equation
(\ref{heat}). If we had taken the same value of $\tau$ for all eigenvalues, the $Z_3$ symmetry would have been broken and the system would evolve to one of the single-cut solutions, even if the starting configuration was very close to the $Z_3$ solution.

\section{Weak coupling free energy}
\label{appendix:weak}

In this appendix we describe the evaluation of the integrals in (\ref{free:energy:general}) for the free-energy of the $U(N)$ pure CS model  at weak coupling. 
  \begin{figure}
\begin{center}
  \includegraphics[width=53mm,angle=0,scale=2.5]{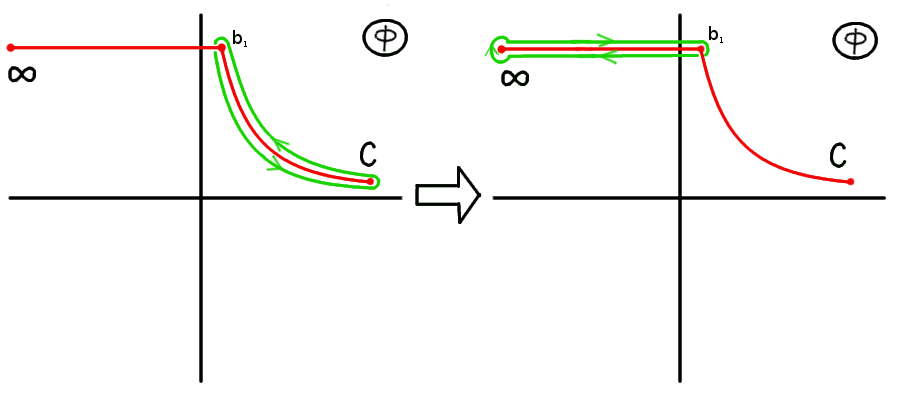}
\end{center}
\caption{Deformation of integration contour in (\ref{free:energy:general}).}
\label{contour}
\end{figure}

The integration contour $\CC$ and its deformation are shown in fig.\,\ref{contour}.  Using the density in (\ref{rhoeq}) with $\kappa=\mu=0$, the first integral in (\ref{free:energy:general}) can be deformed out to infinity and expressed as
\be
\frac{kN\pi}{3}\int_{\mathcal{C}} \phi^3\rho(\phi)d\phi&=&
\frac{N^2\pi^2}{6\tilde\lambda^2}\left(\frac{i}{2\pi}\right)\oint_{\infty} \phi^3\sqrt{\phi^4-\frac{4\tilde\lambda}{\pi}\phi+B}\nn\\
&=&\frac{N^2\pi^2}{6\tilde\lambda^2}\left(\frac{i}{2\pi}\right)\oint_{\infty} \phi^5\left(1-\frac{2\tilde\lambda}{\pi}\phi^{-3}+\frac{B}
{2}\phi^{-4}-\frac{2\tilde\lambda^2}{\pi^2}\phi^{-6}+\dots\right)\nn\\
&=&\frac{N^2}{3}\,,
\ee
which is independent of $B$.  

In the second term we can deform one of integration contours as shown on Fig.\ref{contour},
so that we get
\be\label{doubint}
&&-\frac{N^2}{2}\int_{\mathcal{C}} d\phi d\phi'\rho(\phi)\rho(\phi')\log(\phi-\phi')^2\nn\\
&&=-\frac{2\pi i N^2}{2}\int_{-\infty}^{b_1}\rho(\phi)d\phi\int_{\mathcal{C}_1}\rho(\phi')d\phi'-
\frac{2\pi i N^2}{2}\int_{-\infty}^{b_2}\rho(\phi)d\phi\int_{\mathcal{C}_2}\rho(\phi')d\phi'\nn\\
&&\qquad-\frac{N^2}{4}\oint_{\infty}\rho(\phi)\log\phi^2+\frac{N^2}{2}\oint_{\infty}\phi^{-3}
\rho(\phi)d\phi\,\frac{1}{3}\int_{\mathcal{C}}{\phi'}^3\rho(\phi')d\phi'\,,
\ee
where $\mathcal{C}_1$ and $\mathcal{C}_2$ refer to the two contours of eigenvalues and $b_1$ and
$b_2$ are branch points on those contours.  The first integral on the second line is assumed to 
start and stop at $-\infty$. For any value of $B$ we have that
\be\label{logint}
-\frac{N^2}{4}\oint_{\infty}\rho(\phi)\log\phi^2&=&N^2\lim_{\phi\to-\infty}\left(-\frac{\pi}
{6\tilde\lambda}|\phi|^3-\log|\phi|\right)\nn\\
\frac{N^2}{2}\oint_{\infty}\phi^{-3}\rho(\phi)d\phi\,\frac{1}{3}\int_{\mathcal{C}}{\phi'}^3
\rho(\phi')d\phi'&=&\frac{N^2}{2}\frac{i}{2\tilde\lambda}\oint_{\infty}\frac{\phi^2d\phi}{\phi^3}
\frac{\tilde\lambda}{3\pi}=-\frac{N^2}{6}\,.
\ee
For the other integrals we will consider special cases.  Note that the $\phi'$ integrals give the 
filling fractions for the contours.  For  the $Z_3$ solution  which has $B=0$, we can treat 
the problem as having only one contour since the two contours actually touch at the origin.  We  
then find that
\be\label{rhoint1}
-\frac{2\pi i N^2}{2}\int_{-\infty}^{b_1}\rho(\phi)d\phi&=&\lim_{\phi\to-\infty}N^2\Bigg(\frac{\pi}{6\tilde\lambda}
|\phi|^3+\frac{\pi}{2\tilde\lambda}\frac{2}{3}\frac{\tilde\lambda}{\pi}(1+\log(4)+3\log(\phi))\nn\\
&&\qquad\qquad\qquad\qquad\qquad+
\frac{2\pi}{4\tilde\lambda}\left(-\frac{2\tilde\lambda}{3\pi}\right)\log\left(\frac{4\tilde\lambda
}{\pi}\right)\Bigg)\nn\\
&=&\lim_{\phi\to-\infty}N^2\left(\frac{\pi}{6\tilde\lambda}|\phi|^3+\log|\phi|+\frac{1}{3}-\frac{1}{3}
\log\frac{\tilde\lambda}{\pi}\right)
\ee
For $B=-3\left({\pi}/{\tilde\lambda}\right)^{4/3}$  there is  a single cut with a 
branch point at \mbox{$\phi=\left(-1+i\sqrt{2}\right)\left({\tilde\lambda}/{\pi}\right)^{1/3}$}.  In 
this case we find
\be\label{rhoint2}
-\frac{2\pi i N^2}{2}\int_{-\infty}^{b_1}\rho(\phi)d\phi=\lim_{\phi\to-\infty}N^2\left(\frac{\pi}{6\tilde\lambda}
|\phi|^3+\log|\phi|+\frac{1}{3}-\frac{1}{3}\log\frac{\tilde\lambda}{\pi}+\frac{1}{2}\log 2\right)\,.
\label{integral}
\ee

Finally let us consider the cases with $B^{\pm}=-{3}\left({\pi}/{\tilde\lambda}\right)^{4/3}e^{\pm 2\pi i/3}$,
corresponding to  $Z_3$-rotations of the previous solution.  It is clear that under the change of variables $\phi'=e^{\mp 2\pi i/3}\phi$ that
\be
-\frac{2\pi i N^2}{2}\int_{-\infty}^{b^{\pm}_1}\rho^{\pm}(\phi)d\phi=\lim_{|\phi|\to\infty}-\frac{2\pi i N^2}{2}\int_{-|\phi|e^{\mp 2\pi i/3}}^{b_1}\rho(\phi')\,d\phi'
\ee
where $\rho(\phi)$ and $b_1$ are the same as as in (\ref{integral}). Since the integral only depends on the absolute value of $\phi$ and not its phase, the result is the same as in (\ref{integral}).   Thus the free energy is the same for all three single-cut solutions.

Combining (\ref{logint}) with (\ref{rhoint1}) or (\ref{rhoint2})  we find
\be
&&-\frac{N^2}{2}\int_{\mathcal{C}} d\phi d\phi'\rho(\phi)\rho(\phi')\log(\phi-\phi')^2\nn\\
&&\qquad\qquad=N^2
\left(\frac{1}{6}-\frac{1}{3}\log\frac{\tilde\lambda}{\pi}\right),\qquad\qquad B=0\nn  \\
&&\qquad\qquad=N^2
\left(\frac{1}{6}-\frac{1}{3}\log\frac{\tilde\lambda}{\pi}+\frac{1}{2}\log2\right) , \quad B=-{3}\left(\frac{\pi}{\tilde\lambda}\right)^{4/3}e^{ 2\pi in/3}\,.
\ee
Therefore,
\be
F&=&N^2\left(\frac{1}{2}-\frac{1}{3}\log\frac{\tilde\lambda}{\pi}-C\right)\qquad\qquad \qquad B=0\nn\\
F&=&N^2\left(\frac{1}{2}-\frac{1}{3}\log\frac{\tilde\lambda}{\pi}+\frac{1}{2}\log2-C\right)
\qquad B=-{3}\left(\frac{\pi}{\tilde\lambda}\right)^{4/3}e^{ 2\pi in/3}\,.
\ee

\section{Exact solutions for $m^2=-1/4$}
\label{appendix:exact}

In this appendix we describe  an exact single-cut solution to the full matrix model at the special  value $m^2=-1/4$.
As we emphasized in the main text, while the eigenvalues  lying along the real axis are exponentially close to the exact solution for the strongly coupled pure CS model, the profile  for those along 
the imaginary axis is not as clear because the approximations we assume in the solution break down near the imaginary axis.  Furthermore, the numerical results show that while the eigenvalues are close to the imaginary axis, they appear scattered about it.  Therefore, it is very useful to study any available exact solution to get a better picture of these structures.

At $m^2=-1/4$ the determinant of the partition function 
drastically simplifies and  the pure CS eigenvalue equations reduce to
\begin{eqnarray}
\frac{\pi N}{\tilde\lambda}\phi^2_i&=&2 \pi \sum\limits_{j\neq i}\coth(\pi(\phi_i-\phi_j))\,.
\label{eom2}
\end{eqnarray}
Defining the new variables $u_i=e^{2\pi\phi_i}$, we can put  (\ref{eom2}) into the standard form
\be
\frac{N}{2}V'(u_i)=\sum\limits_{j\neq i}\frac{1}{u_i-u_j}\,,
\ee
where
\be
V(u)=\frac{1}{24\pi^2\tilde\lambda}(\log u)^3+\log u\,.
\ee
In the large $N$ limit, a single cut solution will then have the eigenvalue density
\be\label{rhou}
\rho(u)=-\frac{1}{\pi^2}\sqrt{(a-u)(u-b)}\pintba\frac{du'}{(u-u')\sqrt{(a-u')(u'-b)}}\frac{1}{2}V'(u')
\ee
where the end-points $a$ and $b$ are to be determined.
The resolvent is then defined as
\be
w(u)=\int_b^a \frac{\rho(u')du'}{u-u'}\to \frac{1}{u}\ \mbox{as}\ u\to\infty\,.
\ee
Consistency with the equations of motion and the large $u$ behavior of the resolvent then leads to 
the following constraint equations
\be\label{conds}
\int_b^a\frac{du}{\sqrt{(a-u)(u-b)}}\frac{1}{2}V'(u)&=&0\nn\\
\frac{1}{\pi}\int_b^a\frac{du}{\sqrt{(a-u)(u-b)}}\frac{1}{2}V'(u)u&=&1\,,
\ee
giving us two complex equations for $a$ and $b$, and in principle making them determinable.  If we
define $U(u)=V(u)-\log u$, then the equations take the more symmetric form
\be\label{abcond}
\frac{\sqrt{ab}}{\pi}\int_b^a\frac{du}{\sqrt{(a-u)(u-b)}}U'(u)&=&-1\nn\\
\frac{1}{\pi}\int_b^a\frac{du}{\sqrt{(a-u)(u-b)}}U'(u)u&=&+1\,,
\ee

\begin{figure}
\begin{center}
  \includegraphics[width=53mm,angle=0,scale=2.5]{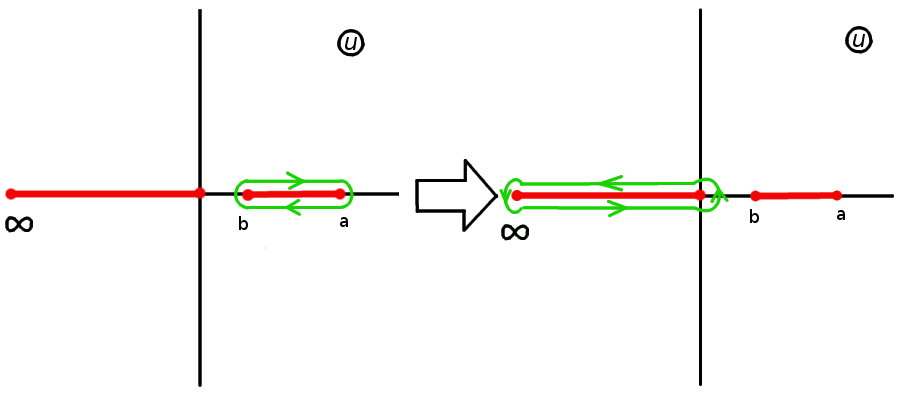}
\end{center}
\caption{Deformation of the integration contour in (\ref{abcond}).}
\label{contour:2}
\end{figure}

The integrals can be done by first assuming that $a>b>0$ and then analytically continuing into the
complex plane.  In the first integral, by deforming the contour as demonstrated in Fig. \ref{contour:2}, we can show that
\be
\int_b^a\frac{(\log u)^2du}{u\sqrt{(a-u)(u-b)}}=\frac{1}{\sqrt{ab}}\left(\pi(\log(-u))^2\Bigg|_{{u\to0_-}}
\!\!\!-\frac{\pi^3}{3}\right)-2\pi\int_{-\infty}^{0_-}\frac{\log(-u)du}{u\sqrt{(a-u)(b-u)}}\,.\nonumber\\
\ee
Integrating by parts, the integral on the rhs becomes
\be\label{refeq1}
\frac{4\pi}{\sqrt{ab}}\log(-u)\,\mbox{arctanh}\left(\sqrt{\frac{b}{a}}\sqrt{\frac{a-u}{b-u}}\right)
\Bigg|_{-\infty}^{0_-}-\frac{4\pi}{\sqrt{ab}}\int_{-\infty}^{0_-}\frac{du}{u}\mbox{arctanh}
\left(\sqrt{\frac{b}{a}}\sqrt{\frac{a-u}{b-u}}\right)\,.
\ee
Letting $d=\sqrt{\frac{a}{b}}$ and defining $y=\sqrt{\frac{a-u}{b-u}}$, the last integral can be written as
\be
-\frac{4\pi}{\sqrt{ab}}\int_{-\infty}^{0_-}\frac{du}{u}\mbox{arctanh}\left(\sqrt{\frac{b}{a}}
\sqrt{\frac{a-u}{b-u}}\right)=\frac{2\pi}{\sqrt{ab}}(d^2-1)\int_1^d\frac{2y dy}{(d^2-y^2)(y^2-1)}\log\frac{d+y}{d-y}\,.\nonumber\\
\ee

This  integral is solvable, where  we find
\be
&&\frac{2\pi}{\sqrt{ab}}(d^2-1)\int_1^d\frac{2y\, dy}{(d^2-y^2)(y^2-1)}\log\frac{d+y}{d-y}=\nn\\
&&\qquad\qquad\frac{2\pi}{\sqrt{ab}}
\Bigg(\frac{\pi^2}{6}+\frac{1}{2}(\log(2d))^2-(\log(d-1))^2+(\log(d+1))^2-\log\frac{d+1}{d-1}\log( 2d)\nn\\
&&\qquad +\Li_2\left(-\frac{2}{d-1}\right)-\Li_2\left(\frac{d-1}{2d}\right)+\Li_2\left(\frac{d-1}{d+1}\right)
+\Li_2\left(\frac{d+1}{2d}\right)\nn\\
&&\qquad+\frac{1}{2}(\log(d-y))^2-\log(2d)\log(d-y)\Bigg|_{y\to d}-\log\frac{d+1}{d-1}\log(y-1)\Bigg|_{y\to1}
\Bigg)\,.
\ee
Combining all terms, the divergences cancel and using several dilogarithm identities, we find 
\be
\frac{\sqrt{ab}}{\pi}\int_b^a\frac{du}{\sqrt{(a\!-\!u)(u\!-\!b)}}U'(u)=\frac{1}{4\pi^2\tilde\lambda}
\left[\frac{1}{2}\left(\log(b)+2\log\frac{2d}{d\!+\!1}\right)^2+\Li_2\left(\left(\frac{d\!-\!1}{d\!+\!1}\right)^2\right)\right]\,.\nonumber\\
\label{refeq2}
\ee

Using similar techniques, one can also show that 
\be
\frac{1}{\pi}\int_b^a\frac{du}{\sqrt{(a\!-\!u)(u\!-\!b)}}U'(u)u=\frac{1}{4\pi^2\tilde\lambda}
\left[\frac{1}{2}\left(\log(a)-2\log\frac{2d}{d\!+\!1}\right)^2+\Li_2\left(\left(\frac{d\!-\!1}{d\!+\!1}\right)^2\right)\right]\,.\nn\\
\ee
Hence, the conditions in (\ref{abcond}) can be reexpressed as
\be\label{abeq}
\frac{1}{2}\left(\log(b)+2\log\frac{2d}{d+1}\right)^2+\Li_2\left(\left(\frac{d-1}{d+1}\right)^2\right)&=&-4\pi^2\tilde\lambda\nn\\
\frac{1}{2}\left(\log(a)-2\log\frac{2d}{d+1}\right)^2+\Li_2\left(\left(\frac{d-1}{d+1}\right)^2\right)
&=&+4\pi^2\tilde\lambda\,.
\ee
These equations can be rewritten in terms of the $\phi$ variables as
\be\label{endpeqs}
\left(\frac{4\pi^2\tilde\lambda}{Z}\right)^2+Z^2+2\,\Li_2\left(1-e^{-Z}))\right)&=&0\nn\\
\phi_{max}+\phi_{min}&=&\frac{4\pi\tilde\lambda}{Z}
\ee
where $a=e^{2\pi\phi_{max}}$, $b=e^{2\pi\phi_{min}}$, and $Z=2\log(\cosh\frac{\pi}{2}(\phi_{max}-\phi_{min}))$

We can also derive an expression for the eigenvalue density $\rho(u)$.  Using the identity
\be
\frac{1}{u-u'}=\frac{u'}{u}\left(\frac{1}{u-u'}+\frac{1}{u'}\right)
\ee
and the first equation in (\ref{conds}),  (\ref{rhou}) can be rewritten as
\be
\rho(u)=-\frac{1}{\pi^2}\frac{\sqrt{(a-u)(u-b)}}{u}\pintba\frac{du'}{(u-u')\sqrt{(a-u')(u'-b)}}\frac{1}{2}V'(u')u'\,.
\ee
The constant piece in $V'(u')u'$ does not contribute to the integral. Deforming the contour to encircle the log branch cut, we then have
\be\label{rhou2}
\rho(u)=\frac{1}{8\pi^3\tilde\lambda}\frac{\sqrt{(a-u)(u-b)}}{u}\int_{-\infty}^0\frac{\log (-u')du'}{(u-u')\sqrt{(a-u')(b-u')}}\,.
\ee
Integrating by parts as in (\ref{refeq1}) and using the same substitutions of variables as in (\ref{refeq1})-(\ref{refeq2}), we find
\be\label{rhof}
\rho(u)&=&\frac{i}{16\pi^3\tilde\lambda}\,\frac{1}{u}\Bigg[4\,\Li_2\left(-\frac{1-y}{1+y}\right)+4\,\Li_2\left(-\frac{1+y}{d-y}\right)-4\,\Li_2\left(-\frac{1-y}{d+y}\right)-4\,\Li_2\left(-\frac{d+y}{d-y}\right)\nn\\
&&\qquad\qquad\quad+2\log b\left(\log\frac{d+y}{d-y}-\log\frac{1+y}{1-y}\right)+\left(\log\frac{d+y}{d-y}-\log\frac{1+y}{1-y}\right)^2\nn\\
&&\qquad\qquad\quad-2\left(\log\frac{d-y}{1-y}\right)^2+2\left(\log\frac{d-y}{1+y}\right)^2-2\left(\log\frac{d+y}{d-y}\right)^2\Bigg]\,.
\ee
The endpoints are at $y=0,\infty$, with the distribution crossing over branch cuts from the logs and dilogarithms.

We first  check these results for $\tl\ll1$.  In this limit the  end points in (\ref{rhou}) approach \mbox{$a=b=1$}.    
 Expanding about $d=1$ we find 
\be
\rho(\phi)=2\pi u\rho(u)&\approx&\frac{i}{8\pi^2\tilde\lambda}\frac{y(d-1)}{y^2-1}\left(2\log b+(d-1)-2\,\frac{d-1}{y^2-1}\right)\nn\\
&\approx&\frac{1}{2\tilde\lambda}\left(\phi+\frac{1}{2}(\phi_{max}+\phi_{min})\right)\sqrt{(\phi_{max}-\phi)(\phi-\phi_{min})}\,,
\ee
which agrees with the density extracted from (\ref{weq}).

Let us now analyze our results in the limit of large $\tilde\lambda$.  In this limit we assume that
$|d|\gg1$, justifying this afterwards.  In this case the equations in (\ref{abeq}) reduce to
\be\label{abres}
\log(b)\approx 2\pi i \sqrt{2(\tilde\lambda+1/24)}-2\log2\qquad\log(a)\approx 2\pi  
\sqrt{2(\tilde\lambda-1/24)}+2\log2\,.
\ee
In terms of the $\phi$ variable this translates to
\be\label{phires}
\phi_{min}\approx  i \sqrt{2(\tilde\lambda+1/24)}-\frac{1}{\pi}\log2\qquad\phi_{max}\approx  
\sqrt{2(\tilde\lambda-1/24)}+\frac{1}{\pi}\log2\,.
\ee
Note that this is consistent to leading order in $\tilde\lambda$ with the endpoints in (\ref{phipm}).  We can also see that with these solutions
\be
|d|=\bigg|\sqrt{\frac{a}{b}}\bigg|\approx 4\exp\left(\pi\sqrt{2(\tilde\lambda-1/24)}\right)\,,
\ee
showing that our approximations are accurate up to exponentially 
small corrections.  

For large $\tl$ we expect half the eigenvalues to extend along the positive real direction.  In this  region of the complex $\phi$ plane we
have $|d|\gg|y|\gg 1$, except very close to the endpoint where $y\to0$.  Away from this endpoint the density in (\ref{rhof})
is approximately
\be
\rho(\phi)=2\pi u\rho(u)&\approx&\frac{i}{8\pi^2\tilde\lambda}\left(4\left(\frac{\pi^2}{6}+0-0-\left(-\frac{\pi^2}
{12}\right)\right)+(-\pi i)^2\right.\nn\\&&\left.-2\pi i\log b-2\log\left(-\frac{d^2}{y^2}\right)\pi i\right)
\approx\frac{1}{2\tilde\lambda}\phi\,,
\ee
which agrees with (\ref{strong:density}).

\begin{figure}
\begin{center}
  \subfigure[]{\label{paths:a}\includegraphics[width=36.5mm,angle=0,scale=1]{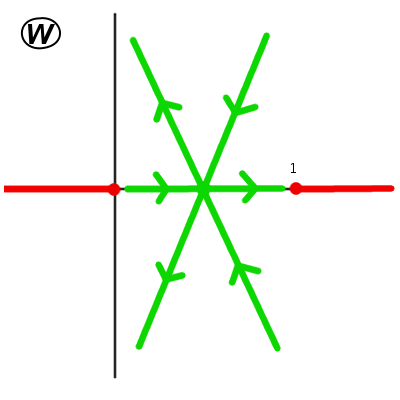}}
  \subfigure[]{\label{paths:b}\includegraphics[width=36.5mm,angle=0,scale=1]{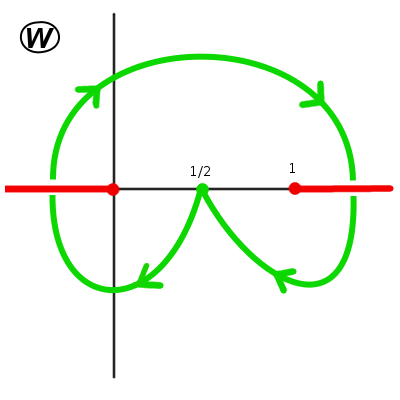}}
   \subfigure[]{\label{paths:c}\includegraphics[width=36.5mm,angle=0,scale=1]{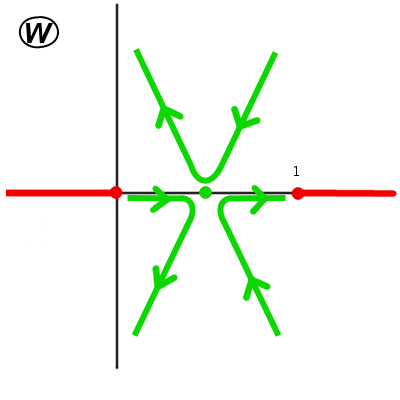}}
   \subfigure[]{\label{paths:d}\includegraphics[width=36.5mm,angle=0,scale=1]{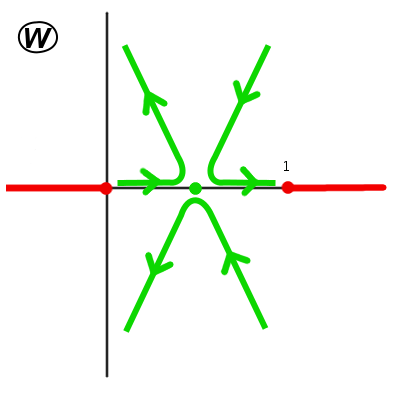}}
\end{center}
\caption{Path of the cut in $w$-plane.}
\label{paths}
\end{figure}

The analysis along the imaginary axis is trickier.  In this case we have that $|y|,|d|\gg1$, but which of these is
bigger depends on the position along the distribution.  It is convenient to define $z=d/y=\sqrt{\frac{1-u/b}{1-u/a}}$.  The density can then be well-approximated by
\be
\rho(\phi)&\approx& \frac{i}{8\pi^2\tilde\lambda}\Bigg(\left(2\log b+\log(4(1-z^2))\right)\log\frac{1+z}{1-z}\nn\\
&&\qquad\qquad+2\Li_2\left(\frac{1+z}{2}\right)-2\Li_2\left(\frac{1-z}{2}\right)\Bigg)\,,
\ee
up to exponentially small corrections.  It is obvious that this is an odd function of $z$.
Furthermore, this can be integrated to give the relatively simple form
\be
n(w)&=&\int_{\phi_{min}}^\phi\rho(\phi)d\phi\nn\\
&=&\frac{i}{16\pi^3\tilde\lambda}\Bigg(\frac{1}{3}(\log(1-w)-\log w)[\pi^2-(\log w)^2-4\log w\log(1-w)-(\log(1-w))^2]
\nn\\
&&\qquad\qquad+4\,\Li_3(w)-4\,\Li_3(1-w)\nn\\
&&\qquad
+\log(4b)\Big((\log w)^2-(\log(1-w))^2+2\Li_2(w)-2\Li_2(1-w)\Big)\Bigg)\,,\nn\\
\ee
where $w=\frac{1+z}{2}$.  $n(w)$ has cuts extending from 1 to $+\infty$ and from $0$ to $-\infty$, and the allowed
$w$ on the eigenvalue path are chosen so that $n(w)$ is positive real.  The eigenvalue path follows a contour that
alternates crossing  the negative branch cut from the bottom and the positive branch cut from the top (see for
example fig.\,\ref{paths:b}).  If we follow 
a path such that we cross each cut $m$ times and return to the same value of $w$, then $n(w)$ shifts by
\be\label{nww}
n(w)\to n(w)-\frac{1}{\tilde\lambda}\left(m^2+\frac{m\,i}{2\pi}\log(4b(w-w^2))\right)\,,
\ee
where $w$ is evaluated on the principle sheet.  In terms of $\phi$ we can rewrite the shift as
\be\label{nw}
n(w)\to n(w)+\frac{m}{\tilde\lambda}\left(m-{i\,}\phi\right)\,,
\ee
where under the transformation $\phi\to\phi-2im$.

We can now argue using (\ref{nww}) that a single contour  is not a viable solution for large $\tilde\lambda$. 
If we are in the principle branch near the beginning of the contour then $n(w)$ is approximately
\be\label{nwseries}
n(w)=-\frac{4\,i\,\log(4b)}{3\tilde\lambda\pi^3}(w-1/2)^3+{\rm O}((w-1/2)^5)\,.
\ee
The three possible choices of contours originating out of the point $w=1/2$ are shown in fig.\,\ref{paths:a}.  
Along these
contours $n(w)$ is positive and increasing.  One of the contours heads directly to the branch point at $w=1$, 
which corresponds to the undesired behavior of $\mbox{Re}(\phi)$ going to negative infinity.  

Instead we should choose the contour that leads to the path shown in  fig.\,\ref{paths:b}
as we move away from $w=1/2$.  In this  case the 
contour will cross both branch cuts and start heading toward the $w=1/2$ point, but now in the $m=1$ branch.
Near this point we should still insist that $n(w)$ is positive and increasing. Using the expansion in 
(\ref{nwseries}) and the shift in (\ref{nww}) we find that
\be\label{nwm=1}
n(w)&\approx &-\frac{4\,i\,\log(4b)}{3\tilde\lambda\pi^3}(w-1/2)^3-\frac{i}{2\pi\tilde\lambda}\log(b)-\frac{1}{\tl}
+\frac{2 i}{\pi\tl}(w-1/2)^2\nn\\
&\approx&\frac{8\sqrt{2}}{3\sqrt{\tilde\lambda}\pi^2}(w-1/2)^3+\frac{\sqrt{2}}{\sqrt{\tilde\lambda}}+\frac{i\log(2)}
{\pi\tilde\lambda}-\frac{1}{\tl}
+\frac{2 i}{\pi\tl}(w-1/2)^2\,,
\ee
with $w$  chosen so that the imaginary part in (\ref{nwm=1}) is zero. Fig.\ref{paths:c} shows 
the contours near 
$w=1/2$ that satisfy this condition.  As can be seen, the contour that heads toward $w=1/2$ on Fig.\ref{paths:c}, 
makes a sharp turn near $w=1/2$ and heads toward the $w=1$ branch point, signifying the breakdown of the 
single cut solution.

This last analysis assumes that $\tilde\lambda$ is real.  If instead we allow for a small imaginary part, 
then we can change the behavior in Fig.\ref{paths:c}.  For example, let $\tilde\lambda=\rho e^{i\theta}$ where $\theta$ 
is assumed to be a small angle.  Then for the $m^{\rm th}$ branch (\ref{nwm=1}) becomes
\be\label{nwm=m}
n(w)\approx\frac{8\sqrt{2}}{3\sqrt{\rho}\pi^2}(w-1/2)^3+\frac{m}{\rho}(\sqrt{2\rho}-m)-i\frac{m}{2\rho}
\theta(\sqrt{2\rho}-2m)+\frac{i\log(2)}{\pi\rho}+\frac{2 i m}{\pi\rho}(w-1/2)^2\,.\nn\\
\ee
If we then have $\theta>\frac{2\log(2)}{\pi(\sqrt{2\rho}-2m)}$ then the contour will behave like fig.\,\ref{paths:d}, at least for small enough $m$, allowing the contour to continue onto the next branch.  However, as $m$ 
approaches $\sqrt{\rho/2}$ then the approximation starts breaking down.

\begin{figure}
\begin{center}
  \subfigure[$\tilde\lambda=0.8$]{\label{cut:break1}\includegraphics[width=40mm,angle=0,scale=1.6]{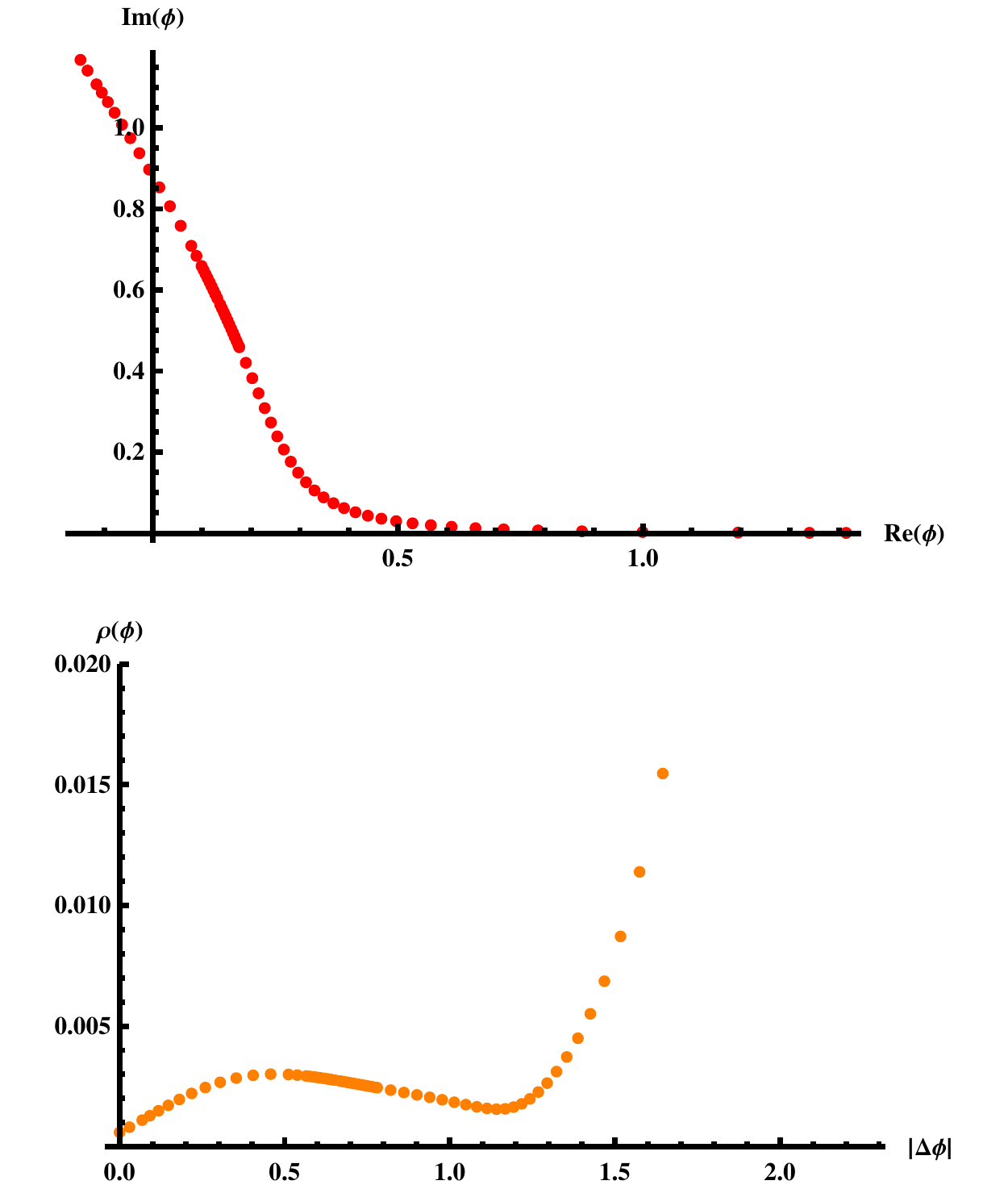}}
  \subfigure[$\tilde\lambda=0.976$]{\label{cut:break2}\includegraphics[width=40mm,angle=0,scale=1.6]{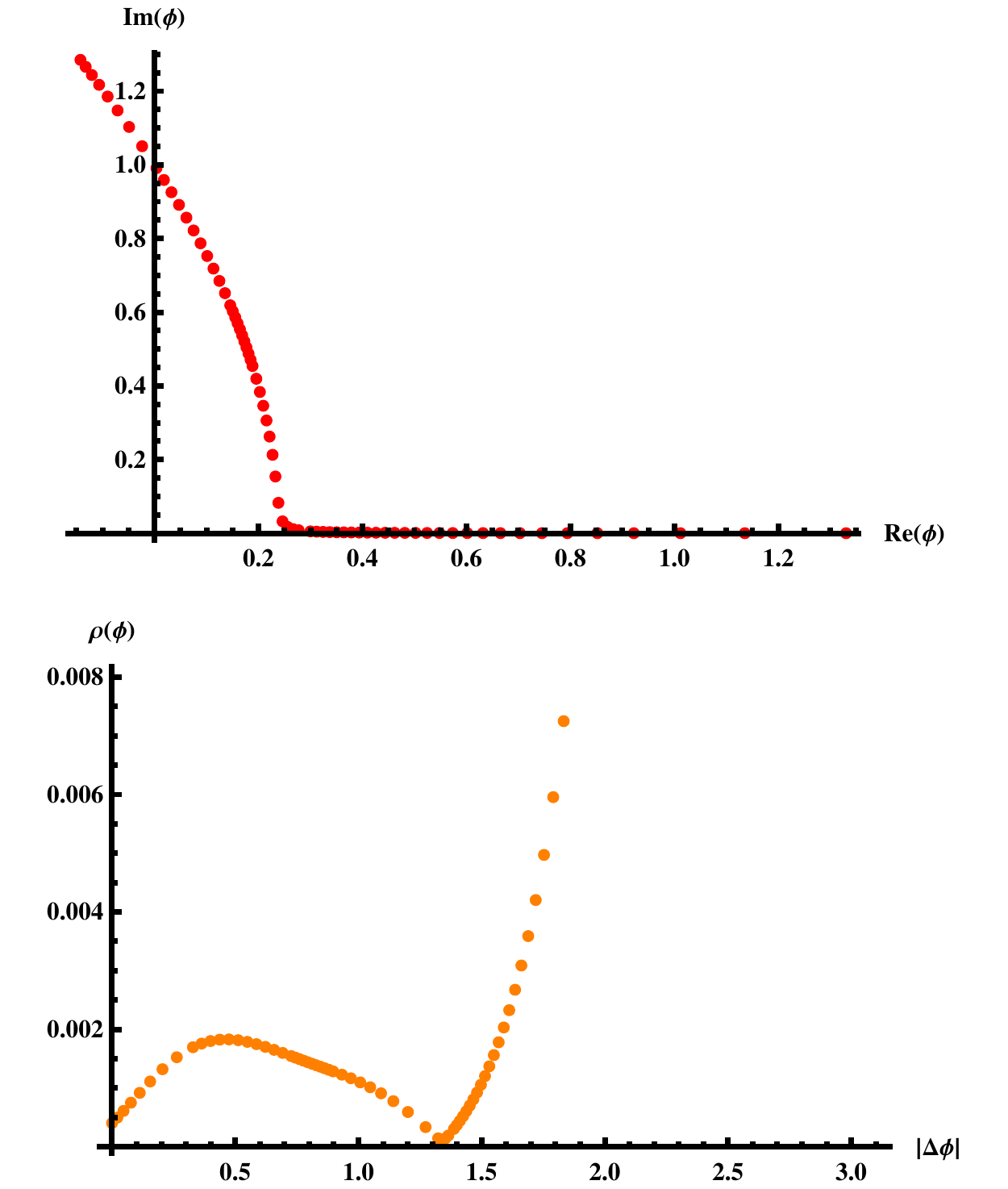}}
\end{center}
\caption{Cut behavior for different $\tilde\lambda$}
\label{cut:break}
\label{paths2}
\end{figure}

Our interpretation of these results is that for large real $\tilde\lambda$ the contour likely splits into 
order $\sqrt{2\tilde\lambda}/2$ separate contours, with each contour roughly between the $w=1/2$ points of successive
branches.  This should at least be true for relatively small values of $m$.  Since the densities are higher for
the small values of $m$, these branches will dominate over the larger values.  If we then give a small imaginary 
part to $\tilde\lambda$, then we expect the contours in the smaller $m$ regions to join together.  The number of 
joined contours will increase as we increase $\theta$.

This also suggests that for real $\tilde\lambda$  there will be a succession of phase transitions 
as $\tilde\lambda$ is increased, with a transition every time $\sqrt{2\tilde\lambda}$ increases by $2$.  Numerically we have found that starting at weak coupling, the single contour degenerates 
at $\tilde\lambda=\tl_c\approx .976$. The behavior of the cut is shown in fig.\,\ref{cut:break}.
As we see from fig.\,\ref{cut:break2}, as $\tilde\lambda\to\tilde\lambda_c$ the density $\rho(\phi)$ goes to zero in the middle of the cut, but still close to the real line.  The cut then breaks in two
above $\tl_c$.  As we increase $\tilde\lambda$ above $\tl_c$ we expect the contour to split every time $\sqrt{2\tilde\lambda}$ increases by $2$

We can also investigate the single-cut solution where the end points are symmetric about the real axis.  To
this end we  let $a=q\, e^{i\theta}$, $b=q\, e^{-i\theta}$, and so $d=e^{i\theta}$.  Subtracting
the first equation in (\ref{abeq}) from the second we arrive at the relation
\be\label{absym}
\log q\log\left(\cos\frac{\theta}{2}\right)=2\pi^2\tilde\lambda\,.
\ee
This shows that $q<1$, thus in the $\phi$ plane the real part of the end-points is less than zero.
We also see that the rhs of (\ref{absym}) is positive real and so $\theta<\pi$,  hence  the end-points
are in the strip $-\pi<\mbox{Im}(\phi)<\pi$.  Using (\ref{absym}) and (\ref{abeq}) we can express the coupling entirely in terms of $\theta$,
\be
\tilde\lambda=-\frac{1}{\pi^2}\log\left(\cos\frac{\theta}{2}\right)\left[-\left(\log\left(\cos\frac{\theta}{2}\right)\right)^2-\frac{1}{2}\,\Li_2\left(-\tan^2\frac{\theta}{2}
\right)\right]^{1/2}\,,
\label{symstrong1}
\ee
and thus $q$ in terms of $\theta$,
\be
\log q=-2\left[-\left(\log\left(\cos\frac{\theta}{2}\right)\right)^2-\frac{1}{2}\,\Li_2\left(-\tan^2\frac{\theta}{2}
\right)\right]^{1/2}\,.
\label{symstrong2}
\ee
In the large $\tilde\lambda$ limit one finds that $\theta\to\pi$ and  $\log q\to -\frac{\pi}{\sqrt{3}}$ up 
to exponentially small corrections. Therefore $a\to b$ and  the endpoints of the cut approach
$\phi=-\frac{1}{2\sqrt{3}}\pm\frac{i}{2}$.  

  To compute $\rho(u)$ we  move slightly away from the limiting values and set $\theta=\pi-\epsilon$ in order to avoid potential divergences.   The equations
  (\ref{symstrong1}) and (\ref{symstrong2}) then become
\be
\tilde\lambda \approx -\frac{1}{2\sqrt{3}\pi}\log\frac{\epsilon}{2}\,,~~~~
\log q \approx -\frac{\pi}{\sqrt{3}}-\frac{\sqrt{3}}{2\pi}\epsilon^2 \log\epsilon\,.
\ee
We  can use  the first equation to rewrite the second as
\be
\log q \approx -\frac{\pi}{\sqrt{3}} + 3\,\tilde\lambda\,\epsilon^2 \,.
\label{logq}
\ee
The endpoints of the cut then take the form
\be
\phi_{max}=-\frac{1}{2\sqrt{3}}+\frac{i}{2}-\frac{i\epsilon}{2\pi}\,,\hspace{8mm}
\phi_{min}=-\frac{1}{2\sqrt{3}}-\frac{i}{2}+\frac{i\epsilon}{2\pi}\,.
\label{endpoints:symmetric}
\ee
Substituting these values into (\ref{rhof}) we obtain
\be
\rho(u)=-\frac{1}{2\pi u}\left(i+\frac{2\sqrt{3}\epsilon}{\pi}\right)\,,
\ee
or in terms of  $\phi$,
\be
\rho(\phi)=2\pi u \rho(u)= -i -\frac{2\sqrt{3}~ \epsilon}{\pi}\,.
\label{rhophi:symm}
\ee

\begin{figure}
\begin{center}
  \includegraphics[width=53mm,angle=0,scale=1.5,natwidth=400,natheight=400]{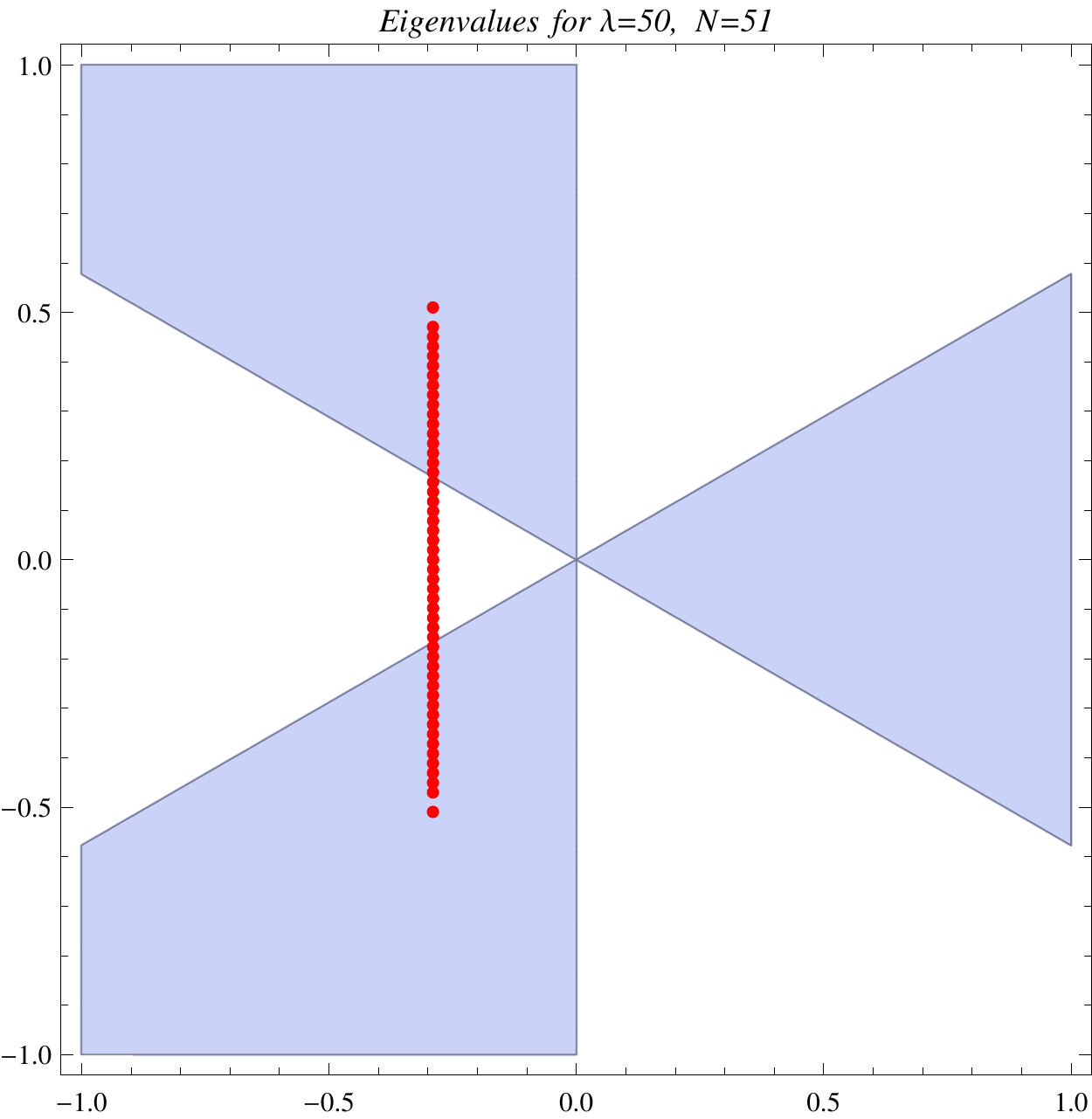}\hspace{25mm}
\end{center}
\caption{Numerical solution symmetric with respect to the real axis for $\tilde\lambda=50$.}
\label{strong:symm:pic}
\end{figure}

We  determine the eigenvalue cut between the endpoints by setting
 $$n(\phi)\equiv\int\limits_
{\phi_{min}}^{\phi}\rho(\phi)~d\phi=-i\left(\phi+\frac{1}{2\sqrt{3}}+\frac{i}{2}\right)$$
 to be positive real. Clearly this is true if we choose the cut to be parallel to the imaginary axis such that $\mbox{Re}(\phi) = -\frac{1}{2\sqrt{3}} $. In  
fig.\,\ref{strong:symm:pic} we show the numerical solution for $\tl=50$, which confirms this behavior. 
Since this cut is of finite extent in the infinite $\tl$ limit, the free-energy can only scale as $N^{2}$ and not $N^{5/2}$.
 
 It is interesting to determine the  behavior of the Wilson loop for this solution since it connects to the weakly coupled real  single-cut solution in (\ref{wilson:asymp:weak:Z2}).
 Using the  eigenvalue density in (\ref{rhophi:symm}) and the endpoint positions  in (\ref{endpoints:symmetric}), we find that (\ref{wilson:loop:mm}) gives for the log of the fundamental Wilson loop,
  \be
\log \langle W\rangle^+_{str.symm.} = -2\sqrt{3} \pi\tilde\lambda
\ee
 This result parallels the decreasing behavior in (\ref{wilson:asymp:weak:Z2}). For other values of $m^2$, including $m^2=0$, we can show numerically that their Wilson loops are also decreasing with $\tl$.

\vfill\eject

\bibliographystyle{JHEP}
\bibliography{nnhalfcs}  
 
\end{document}